\documentclass[aps,reprint,twocolumn,showpacs,preprintnumbers,amsmath,amssymb,nofootinbib,superscriptaddress,showkeys]{revtex4-1}

\usepackage{epsfig}

\begin{document}

\title{Perturbative Renormalizability of Chiral Two Pion Exchange \\
in Nucleon-Nucleon Scattering: P- and D-waves}
	\author{M. Pav\'on Valderrama}\email{m.pavon.valderrama@ific.uv.es} 
\affiliation{Instituto de F\'{\i}sica Corpuscular (IFIC), Centro Mixto CSIC-Universidad de Valencia, Institutos de Investigaci\'on de Paterna, Aptd. 22085, E-46071 Valencia, Spain}

\date{\today}

\begin{abstract} 
\rule{0ex}{3ex} 
We study the perturbative renormalizability of chiral two pion exchange
in nucleon-nucleon scattering for p- and d-waves
within the effective field theory approach.
The one pion exchange potential is fully iterated at the leading order
in the expansion, a choice generating a consistent and
well-defined power counting that we explore in detail.
The results show that perturbative chiral two pion exchange reproduces
the data up to a center-of-mass momentum of $k_{\rm cm} \sim 300\,\rm MeV$
at next-to-next-to-leading order and that the effective field theory expansion
convergences up to $k_{\rm cm} \sim 350\,\rm MeV$.
\end{abstract}

\pacs{03.65.Nk,11.10.Gh,13.75.Cs,21.30.-x,21.45.Bc}
\keywords{Potential Scattering, Renormalization, Nuclear Forces, Two-Body System}

\maketitle

\section{Introduction}

The nature and derivation of the nucleon-nucleon interaction
is probably the central problem of nuclear physics.
The modern point of view is that any serious theoretical formulation
of the nuclear force should be grounded on quantum chromodynamics (QCD),
the fundamental theory of the strong interactions.
However direct calculations on the lattice are still unavailable
at the physical pion mass,
despite promising results which are beginning to appear
in the two~\cite{Beane:2006mx,Ishii:2006ec} and three~\cite{Doi:2011gq}
nucleon sectors at larger pion masses.
A different, more indirect path is provided by the effective field theory (EFT)
formulation of the nuclear forces~\cite{Beane:2000fx,Bedaque:2002mn,Epelbaum:2005pn,Epelbaum:2008ga,Machleidt:2011zz},
in which the pion exchanges are constrained
by the requirement of broken chiral symmetry,
the main low energy manifestation of QCD.
This approach exploits the well-known separation of scales appearing
in the two-nucleon system with the purpose of generating a
systematic and model independent low energy expansion of
the nuclear potential and the scattering amplitudes.
In addition, nuclear EFT may potentially bridge current lattice QCD
calculations with physical results 
by means of  chiral extrapolations.

In the nuclear EFT proposed by Weinberg~\cite{Weinberg:1990rz,Weinberg:1991um}
the nucleon-nucleon potential is expanded as a power series
(i.e. a power counting) in terms of the parameter $Q / \Lambda_0$
\begin{eqnarray}
V_{\rm NN}(r) = V^{(0)}(r) + V^{(2)}(r) +
V^{(3)}(r) + \mathcal{O}\left( \frac{Q^4}{\Lambda_0^4} \right) \, ,
\nonumber \\ 
\label{eq:chiral-pot}
\end{eqnarray}
where $Q$ represents the low-energy scales of the system, usually
the momentum $p$ exchanged between the nucleons and the pion mass $m_{\pi}$,
while $\Lambda_0$ stands for the high-energy scales that
are not explicitly taken into account in the theory,
e.g. the rho meson mass $m_{\rho}$.
The full potential is then iterated into the Schr\"odinger or Lippmann-Schwinger
equation~\cite{Ordonez:1993tn,Ordonez:1995rz,Frederico:1999ps,Eiras:2001hu,Entem:2001cg,Entem:2002sf,Entem:2003ft,Epelbaum:2003gr,Epelbaum:2003xx,Epelbaum:2004fk,Nogga:2005hy,PavonValderrama:2005gu,PavonValderrama:2005wv,PavonValderrama:2005uj,Krebs:2007rh,Higa:2007gz,Valderrama:2008kj,Valderrama:2010fb,Entem:2007jg,Yang:2007hb,Yang:2009kx,Yang:2009pn,Albaladejo:2011bu},
or more recently within the nuclear lattice approach~\cite{Borasoy:2006qn,Borasoy:2007vy,Borasoy:2007vi,Borasoy:2007vk,Epelbaum:2008vj,Epelbaum:2009zsa,Epelbaum:2009pd,Epelbaum:2010xt,Epelbaum:2011md}
~\footnote{It should be noted that lattice EFT implementations of
the Weinberg counting only iterate the order $Q^0$ piece of
the interaction, while the subleading pieces generally enter as perturbations.
The $C_2 (p^2 + p'^2)$ short range operator is an exception as it is promoted
from order $Q^2$ to order $Q^0$ to form an ``improved'' LO interaction.},
in order to obtain theoretical predictions.
Regarding the notation,
order $Q^0$ will also be referred to as leading order (${\rm LO}$),
order $Q^2$ as next-to-leading order (${\rm NLO}$), order $Q^3$ as
next-to-next-to-leading order (${\rm N^2LO}$), and so on.

Among the advantages of the Weinberg prescription we can count that
it observes the non-perturbative nature of the nuclear forces and 
naturally fits into the traditional nuclear physics paradigm of
employing a potential as the basic calculational input.
In particular, the construction of potential descriptions of the nuclear force
is conceptually straightforward, as the procedure only involves the
expansion of the chiral potential up to a given order and
the determination of the free parameters of the theory
by a fit to the available low-energy scattering data.
Eventually, at high enough order, in particular ${\rm N^3LO}$,
this procedure leads to chiral potentials that
are able to reproduce the two-nucleon scattering data
below a laboratory energy of $E_{\rm lab} = 300\,{\rm MeV}$
with a $\chi^2 / d.o.f. \simeq 1$~\cite{Entem:2003ft,Epelbaum:2004fk},
a remarkable degree of success which justifies
the popularity of the Weinberg prescription.

However, the theoretical basis of the Weinberg counting is debatable and,
as we will argue, far from robust.
In particular we can identify two serious problems: (i) the counterterms
included in the Weinberg counting are not enough as to renormalize the
amplitudes (either at ${\rm LO}$~\cite{Eiras:2001hu,Nogga:2005hy,PavonValderrama:2005gu}
or ${\rm NLO}$/${\rm N^2LO}$~\cite{PavonValderrama:2005wv,PavonValderrama:2005uj})
and (ii) it is not clear whether the scattering amplitudes obey
the power counting of the chiral potential as a result of
the full iteration process.
In addition, there is the issue of the chiral inconsistency~\cite{Kaplan:1996xu}
which inspired the Kaplan, Savage and Wise (KSW)
counting~\cite{Kaplan:1998tg,Kaplan:1998we}.
Without these ingredients, renormalizability and power counting,
the Weinberg prescription is merely reduced to a recipe
for constructing nuclear potentials.

The renormalizability problem is a consequence of the appearance of
singular interactions in the chiral expansion of the potential:
the order $\nu$-th contribution to the potential behaves as
$V^{(\nu)}(r) \sim 1/r^{3+\nu}$ for $m_{\pi} r \leq 1$. 
In this regard the theory of non-perturbative renormalizability 
developed in Refs.~\cite{Beane:2000wh,Nogga:2005hy,PavonValderrama:2005gu,PavonValderrama:2005wv,PavonValderrama:2005uj}
states that, if the finite range potential is singular and attractive,
counterterms need to be included for removing the cut-off dependence.
This result implies in particular the non-renormalizability of
the Weinberg counting at any order, as there is always an infinite
number of partial waves for which the chiral potential is singular
and attractive.
Two possible solutions are (i) to treat the chiral
potential, in particular one pion exchange (OPE),
perturbatively in partial waves with sufficiently high angular momentum,
as advocated in Refs.~\cite{Nogga:2005hy,Birse:2005um},
or (ii) to realize that the infinite number of counterterms
is partly an artifact of the partial wave projection that
can be avoided by correlating the short-range operators
of different channels~\cite{Valderrama:2010fb}.
In the present work we will follow the first solution, which
is the simplest and most natural one
from the EFT perspective.

On the contrary, a repulsive singular interaction is completely
insensitive to the value of the counterterms.
This condition leads to unexpected consequences, in particular
the {deuteron ${\rm NLO}$ disaster}~\cite{PavonValderrama:2005wv}:
at ${\rm NLO}$ the chiral potential turns out to be repulsive
in the triplet channel leading to the disappearance of
the deuteron as the cut-off is removed,
in clear disagreement with the experimental evidence.
In addition, the eventual change of sign of the chiral potential
at certain orders prevents the formulation
of a sensible power counting in a fully non-perturbative set-up,
as has been recently stressed by Entem and Machleidt~\cite{Machleidt:2010kb}.

However, the difficulties associated with non-perturbative renormalizability
are not the only reason to avoid the iteration of
chiral two pion exchange (TPE).
A different, but also potentially serious issue is the possible mismatch
between the power counting in the potential and the physical observables.
As is well known, the Weinberg prescription defines power counting
in terms of the ${\rm NN}$ potential, which is not an observable.
The iteration of the potential can alter the relative importance of
each component of the interaction, specially if the cut-off is not soft
enough~\cite{Yang:2009kx,Yang:2009pn,Lepage:1997cs,Epelbaum:2009sd}, 
and thus destroy the original power counting at the level of observables.
In this respect the preliminary results of Ref.~\cite{Valderrama:2010aw}
for the singlet channel indicate that this may be happening
for the typical values of the cut-off employed
in the Weinberg scheme,
calling for a perturbative reanalysis of
the ${\rm N^2LO}$~\cite{Entem:2001cg,Epelbaum:2003gr,Epelbaum:2003xx}
and ${\rm N^3LO}$~\cite{Entem:2003ft,Epelbaum:2004fk} results.

As argued in the previous paragraphs,
the combined requirement of renormalizability and power counting
imposes stringent constraints on the set of theoretically
acceptable EFT formulations of the nuclear force.
At leading order the non-perturbative features of the two-nucleon interaction
requires the iteration of certain pieces of the interaction, minimally
a contact operator in the $^1S_0$ and $^3S_1$
channels~\cite{Kaplan:1998tg,Kaplan:1998we,vanKolck:1998bw},
and optimally the OPE potential in s- and  p-waves (and eventually d-waves),
plus the extra counterterms required to renormalize
the scattering amplitude~\cite{Nogga:2005hy}.
In order to avoid counterterm proliferation, we should be conservative
with respect to what pieces of the chiral potential we iterate.
Finally, to guarantee the existence of a power counting
and prevent the renormalization issues we have mentioned,
the subleading pieces of the interaction should be
perturbative corrections over the ${\rm LO}$ results.
The modified Weinberg scheme of Ref.~\cite{Nogga:2005hy} 
fulfills these conditions and is in addition
a phenomenologically promising approach,
as can be inferred from its lowest order results.
However, as we are dealing with {\it a posteriori} power counting,
the previous is not necessarily the only possible realization
of a consistent and phenomenologically acceptable nuclear EFT,
and there may be other frameworks or variations 
(see Refs.~\cite{Beane:2001bc,Beane:2008bt} for two examples)
that may eventually work as well.

The present work explores the perturbative renormalizability of
chiral TPE at ${\rm NLO}$ and ${\rm N^2LO}$
assuming the non-perturbative treatment of OPE at ${\rm LO}$.
This formulation is to be identified with
the proposal of Nogga, Timmermans and van Kolck~\cite{Nogga:2005hy}
for constructing a consistent power counting for nuclear EFT.
Regarding the formal aspects of the theory,
we elaborate upon
the perturbative renormalization framework of Ref.~\cite{Valderrama:2009ei}
and determine the general conditions that ensure the renormalizability of
the subleading corrections to the ${\rm LO}$ scattering amplitudes.
That is, we deduce the power counting of the contact operators.
As expected the scaling of the counterterms is very similar
to the one previously obtained by Birse~\cite{Birse:2005um}.
As an application of the formalism,
we compute the p- and d-wave phase shifts resulting in a phenomenologically
acceptable description which improves over the Weinberg counting
at the same order~\cite{Epelbaum:2003xx}.

The general approach we follow and the perturbative techniques we employ,
which are based on distorted wave Born approximation (DWBA), 
are equivalent to the momentum space formulation
of Refs.~\cite{Long:2007vp,Long:2011qx}.
In particular, the recent exploration of the $^3P_0$ partial wave
by Long and Yang~\cite{Long:2011qx} leads to conclusions that
are equivalent to the ones obtained in the present work
for the aforesaid wave.
The role of perturbative chiral TPE has also been studied in
the ``deconstruction'' approach of Refs.~\cite{Birse:2007sx,Birse:2010jr,Ipson:2010ah},
which analyzes the scaling of the short range interaction for most of
the uncoupled s-, p- and d-waves within the power counting
derived from Ref.~\cite{Nogga:2005hy}.
In addition deconstruction provides interesting insights
on important aspects of the theory such as the determination
of the expansion parameter of the EFT,
which we will employ in the present work.

The article is structured as follows:
in Sect.~\ref{sec:modified-W} we review the modifications induced by
the requirement of renormalizability to the Weinberg power counting
at leading~\cite{Nogga:2005hy} and subleading~\cite{Valderrama:2009ei} orders.
In Sect.~\ref{sec:pert-reno} we explore in detail the divergences
related to the perturbative treatment of chiral TPE within the DWBA
to determine the power counting of the subleading contact operators
at ${\rm NLO}$ and ${\rm N^2LO}$.
The p- and d-waves phase shifts resulting from this renormalization
scheme are presented in Sect.~\ref{sec:results}.
Finally, we discuss the results and present our conclusion
in Sect.~\ref{sec:conclusions}.
In the appendix \ref{app:LO} we explain the technical details behind
the ${\rm LO}$ non-perturbative renormalization.

\section{Modifications to the Weinberg Power Counting}
\label{sec:modified-W}

In this section we want to explain in an informal manner
why renormalizability requires certain modifications
to the Weinberg counting.
The starting point is the work of Nogga, Timmermans
and van Kolck~\cite{Nogga:2005hy},
which pointed out, by means of a thorough numerical exploration,
that cut-off independence of the leading order nucleon-nucleon
scattering amplitude requires new counterterms
in partial waves for which the ${\rm LO}$ tensor force is attractive.
This precise pattern is naturally explained within the non-perturbative
renormalization framework developed in Refs.~\cite{PavonValderrama:2005gu,PavonValderrama:2005wv,PavonValderrama:2005uj},
which proved that the inclusion of a specific number of counterterms
is a necessary and sufficient condition for the renormalizability
of attractive singular potentials.
Curiously, the failure of the Weinberg scheme is not accidental:
the existence of singular interactions in the EFT expansion of
the chiral potential is just a consequence of power counting itself.
We will illustrate this point in Sect.~\ref{sub:sub-singular}.
For exemplifying the ideas behind the non-perturbative renormalization
of singular potentials and the  ${\rm LO}$ modifications
of the power counting~\cite{Nogga:2005hy,PavonValderrama:2005gu,PavonValderrama:2005wv,PavonValderrama:2005uj},
we will consider the particular case of the $^3P_0$ partial wave
in Sect.~\ref{sub:leading-order}.

Nevertheless, the purpose of the present work is to explore the extension
of the Nogga et al.~\cite{Nogga:2005hy} counting to subleading orders.
As already conjectured in Ref.~\cite{Nogga:2005hy},
the most adequate approach
for formulating a canonical EFT expansion of the amplitudes
is the perturbative treatment of the higher order
corrections to the chiral nuclear potential and the addition
of a certain number of counterterms to guarantee
the renormalizability of the approach.
The power counting for the subleading contact interactions, that is,
the order of the promoted counterterms, was constructed for the
first time by Birse in Ref.~\cite{Birse:2005um}
(see also Ref.~\cite{Birse:2009my} for a more accessible presentation).
More recently, Ref.~\cite{Valderrama:2009ei} proved the viability
of the previous proposal
and found some deviations from the original
power counting of Ref.~\cite{Birse:2005um}.
Incidentally, the subleading power counting deviates from the original Weinberg
prescription even in partial waves for which the leading order counting
remained unaltered in the Nogga et al. proposal~\cite{Nogga:2005hy}.
The best example is provided by the $^1S_0$ partial wave: although
at ${\rm LO}$ both Weinberg and Nogga et al. predict the same number of
counterterms, at ${\rm NLO}$ and ${\rm N^2LO}$ the counterterms
predicted by Weinberg are not enough as to renormalize the
scattering amplitude in this partial wave.
Based on the related discussion of Ref.~\cite{Valderrama:2010aw},
we will explain the power counting of the singlet channel
in Sect.\ref{sub:n2lo}.

\subsection{Singular Potentials}
\label{sub:sub-singular}

The appearance of singular interactions in the chiral potential is
the fundamental reason which requires the power counting
modifications suggested by Nogga et al.~\cite{Nogga:2005hy}.
Curiously, singular potentials are a consequence of
the existence of a power counting
in the potential itself.
The argument for the previous statement is relatively straightforward.
We first consider the scaling of the order $\nu$-th contribution
to the chiral potential, which according to power counting
is given by
\begin{eqnarray}
\label{eq:Vnu-scaling}
V^{(\nu)} \propto \frac{Q^{\nu}}{\Lambda_0^{\nu}} \, ,
\end{eqnarray}
where $Q$ represents the light scales of the system,
like the pion mass or the relative momentum of the nucleons,
while $\Lambda_0$ stands for the heavy scales, for example
the mass of the $\rho$ meson.

It is interesting to notice that the two aforementioned light scales
play a very different role: while the pion mass $m_{\pi}$ is fixed,
the momentum of the nucleons can vary, and in particular it can
take values that are much larger than the pion mass,
that is, $|\vec{q}\,| \gg m_{\pi}$.
In this regime the momentum space representation of $V^{(\nu)}$
fulfills the approximate scaling law
\begin{eqnarray}
V^{(\nu)}(\lambda\,\vec{q}) \propto \lambda^{\nu}\,V^{(\nu)}(\vec{q}) \, ,
\end{eqnarray}
which is just a restatement of Eq.~(\ref{eq:Vnu-scaling}).
After Fourier transforming, this translates into the following
scaling relation in coordinate space
\begin{eqnarray}
V^{(\nu)}(\lambda\,\vec{r}) \propto
\frac{V^{(\nu)}(\vec{r})}{\lambda^{\nu + 3}} \, ,
\end{eqnarray}
which is valid for $m_{\pi} |\vec{r}| \ll 1$.

There are two kinds of solution to the scaling relations induced by the power
counting of the potential: (i) zero range and (ii) finite range solutions.
The zero (or contact) range solutions are constructed from the scaling
properties of the three dimensional Dirac delta, that is
\begin{eqnarray}
\delta (\lambda \vec{r}) = \frac{\delta (\vec{r})}{\lambda^3} \, ,
\end{eqnarray}
which corresponds to a solution of the scaling equation for $\nu = 0$,
yielding
\begin{eqnarray}
V_C^{(0)}(\vec{r}) = C_0 \, \delta(\vec{r}) \, ,
\end{eqnarray}
where the subscript $_C$ indicates the contact range nature of
the contribution.
Higher order contact range solutions can be easily obtained by 
adding derivatives to the Dirac delta, although due to parity
conservation only an even number of derivatives can appear.
That is, new contact terms enter only at at even orders,
$\nu = 2 n$, as is well known.

The finite range solutions of the scaling relations are
fairly straightforward and take the form
\begin{eqnarray}
V^{(\nu)}_F(r) \propto \frac{1}{r^{3+\nu}} \, , 
\end{eqnarray}
where the subscript $_F$ is used to indicate the finite range.
This form corresponds to the observed degree of divergence of
the pion contributions to the chiral potentials~\footnote{
The inclusion of light but static degrees of freedom,
like the $\Delta$ isobar in the small scale expansion~\cite {Hemmert:1997ye},
can alter however the previous scaling.}.
As we will see, the renormalizability problems arise from the non-perturbative
interplay between contact and finite range operators, which modifies their
scaling properties at the level of observables,
thus changing the power counting of the operators.

\subsection{Leading Order Modifications of the Counting}
\label{sub:leading-order}

\begin{figure*}[htt!]
\begin{center}
\epsfig{figure=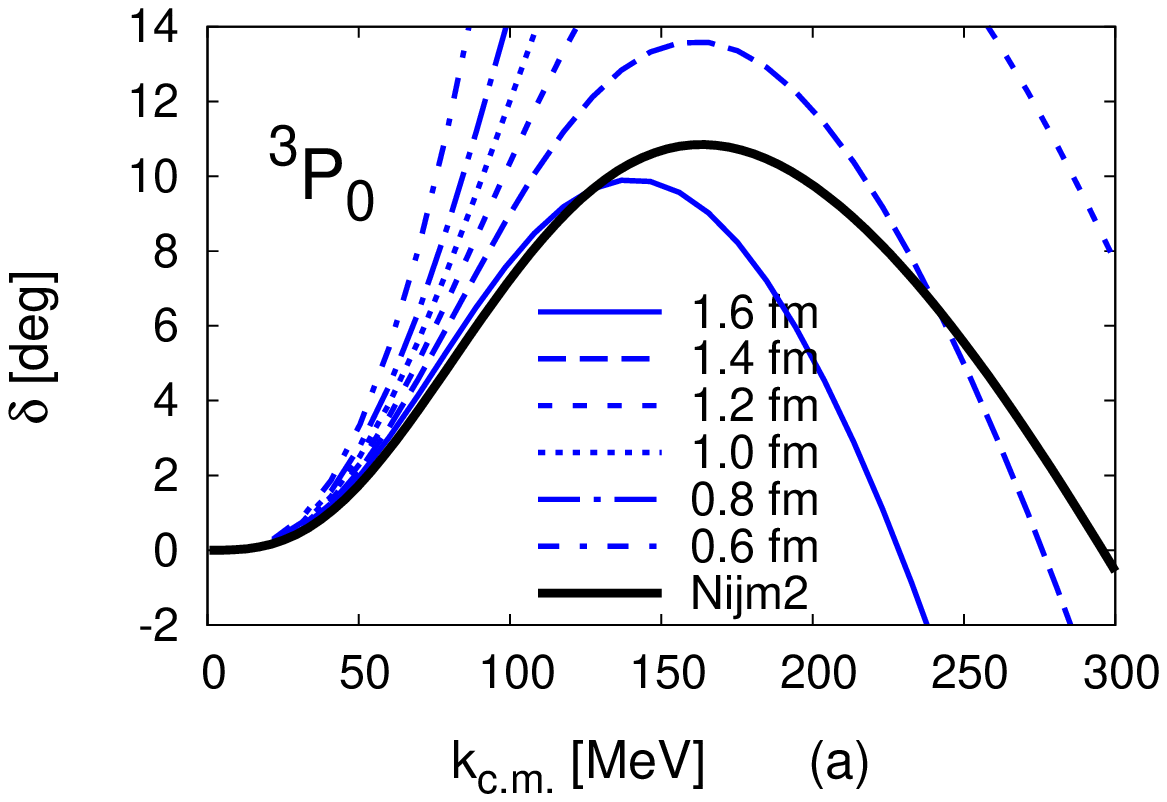,
	height=4.75cm, width=8.0cm}
\epsfig{figure=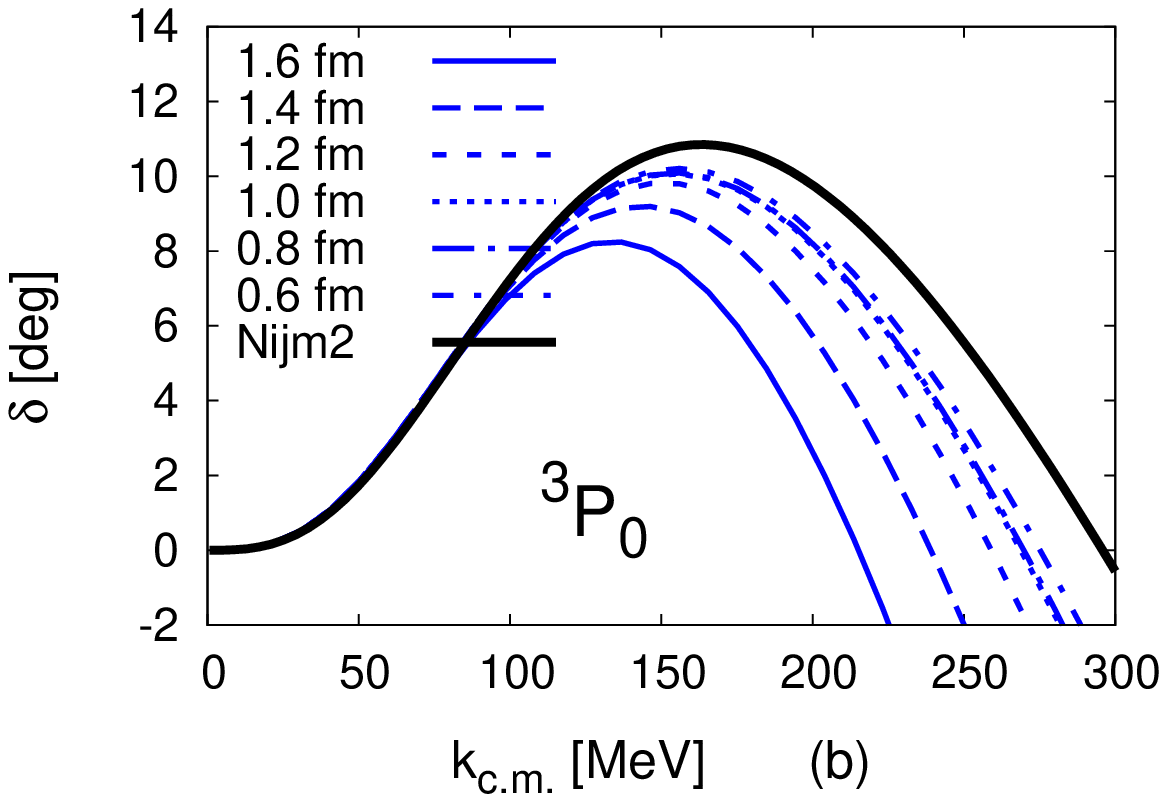, 
	height=4.75cm, width=8.0cm}
\end{center}
\caption{(Color online)
Leading order phase shifts for the $^3P_0$ channel in the Weinberg counting
(left panel) and in the modified Weinberg scheme of Nogga, Timmermans and
van Kolck~\cite{Nogga:2005hy} (right panel).
As can be seen, the $^3P_0$ phases develop a strong cut-off dependence
in the Weinberg counting, which can be eliminated by the promotion of
a counterterm as advocated in Ref.~\cite{Nogga:2005hy}.}
\label{fig:modified-weinberg-leading}
\end{figure*}

Weinberg power counting implicitly assumes that the counterterms appearing
at each order in the chiral expansion are capable of renormalizing
the scattering amplitude.
However, this is not the case, as has been repeatedly discussed 
in the literature~\cite{Kaplan:1996xu,Nogga:2005hy,PavonValderrama:2005wv,PavonValderrama:2005uj}.
In particular, Nogga et al.~\cite{Nogga:2005hy} found numerically
that even at leading order the counterterms of the Weinberg counting
do not render the phase shifts unique:
several counterterms should be added in the $^3P_0$,
$^3P_2-{}^3F_2$ channels and eventually the $^3D_2$ channel,
depending on the cut-off region under consideration
(see also the related comments of Ref.~\cite{Epelbaum:2006pt})~\footnote{
Notice however that the recent application of the ${\rm N/D}$ method
to nuclear EFT~\cite{Albaladejo:2011bu}
does not require the inclusion of an additional counterterm
in the $^3P_0$ or the $^3D_2$ partial waves.}.

The $^3P_0$ channel probably provides the best example to exemplify the
ideas behind the Nogga et al. proposal~\cite{Nogga:2005hy}.
We start by consider this partial wave in the Weinberg counting,
where we can explicitly check the appearance of a strong cut-off dependence
of the $^3P_0$ phase shift at ${\rm LO}$ .
In the Weinberg scheme, the wave function in this channel can be described
by the following Schr\"odinger equation
\begin{eqnarray}
\label{eq:schroe-3P0}
-{u_k^{(0)}}'' + \left[ 2\mu\,V^{(0)}_{^3P_0}(r) +
\frac{2}{r^2} \right]\,u_k(r) = 
k^2 \, u_k^{(0)}(r) \, ,
\end{eqnarray}
where $V^{(0)}_{^3P_0}(r)$ is the order $Q^0$ chiral potential.
As the potential diverges as $1/r^3$ at short distances, we include
a regularization scale $r_c$ which serves as a separation scale
between the unknown short range physics and the known long range
physics.
In particular, we only consider the chiral potential to be valid for $r > r_c$.
For radii below the regularization scale, $r < r_c$,
we do not know how the system can be described,
but it is reasonable to assume that at short enough distances the behaviour
of the wave function will be dominated by the centrifugal barrier, that is,
\begin{eqnarray}
u_k^{(0)}(r) \simeq a \, r^2 \, ,
\end{eqnarray}
for $r < r_c$.
The previous assumption can be effectively translated into a logarithmic
boundary condition for the Schr\"odinger equation at $r = r_c$
\begin{eqnarray}
\label{eq:bc-3P0}
\frac{u_k'(r)}{u_k(r)} \Big|_{r = r_c} \simeq \frac{2}{r_c} \, ,
\end{eqnarray}
corresponding to the radial regulator employed
in Ref.~\cite{PavonValderrama:2005uj}.
By integrating the Schr\"odinger equation from $r = r_c$ to $r \to \infty$
with the previous initial integration condition,
we can extract the phase shifts by matching to the asymptotic form
of the wave function at large distances, which is given by
\begin{eqnarray}
u^{(0)}_k(r) \to
\cot{\delta^{(0)}_{{}^3P_0}}\,\hat{j}_1(k r) - \hat{y}_1(k r) \, ,
\end{eqnarray}
where $\delta^{(0)}_{{}^3P_0}$ is the ${\rm LO}$ $^3P_0$ phase shift,
and $\hat{j}_1(x) = x\,j_1(x)$, $\hat{y}_1(x) = x\,y_1(x)$, with
$j_1(x)$ and $y_1(x)$ the Spherical Bessel functions.

The results for the $^3P_0$ phase shifts in the previous regularization
scheme are depicted in Fig.~(\ref{fig:modified-weinberg-leading}).
The phase shifts, which have been computed for a series of cut-off radii ranging
from $r_c = 0.6\,{\rm fm}$ to $1.6\,{\rm fm}$, show a remarkable
cut-off dependence not anticipated by the Weinberg counting.
The formal reason for this cut-off dependence lies
in the details of the short range ($m_{\pi} r \ll 1$) solution to
the Schr\"odinger equation Eq.~(\ref{eq:schroe-3P0}),
which reads
\begin{eqnarray}
u_k^{(0)}(r) \propto (k a_3)\,\left(\frac{r}{a_3}\right)^{3/4}\,
\sin{\left[ \sqrt{\frac{a_3}{r}} + \varphi \right]} \, ,
\end{eqnarray}
where $a_3$ is a length scale related to the strength of
the tensor force in this channel.
As can be seen, the determination of the wave function near the origin
requires the determination of the phase $\varphi$.
The boundary condition, Eq.~(\ref{eq:bc-3P0}), just chooses a different,
non-converging value of $\varphi$ for each $r_c$, and therefore does not
yield unique results.
In particular, there is not a well defined limit of $\varphi$ for $r_c \to 0$:
the value of $\varphi$ simply oscillates faster and faster
on the way to the origin.

From the EFT point of view, the previous cut-off dependence means that
the a priori estimation of the strength of the contact term for the
$^3P_0$ channel was not correct.
Otherwise, the cut-off dependence would have not appeared until radii of
the order of the breakdown scale of the theory $\Lambda_0 r_c \sim 1$.
However, as can be seen in the left panel of
Fig.~(\ref{fig:modified-weinberg-leading}),
the cut-off dependence manifest for scales of the order of
$m_{\pi} r_c \sim 1$.
This indicates the necessity of including a new short range operator
in this channel at ${\rm LO}$.

By including the new counterterm, the previous uncontrolled cut-off dependence
of the $^3P_0$ phase shift disappears,
as can be appreciated in the right panel of
Fig.~(\ref{fig:modified-weinberg-leading}).
We have employed the regularization scheme of Refs.~\cite{PavonValderrama:2005gu,PavonValderrama:2005wv,PavonValderrama:2005uj},
in which to unambiguously determine the ${\rm LO}$ phase shifts
in the $^3P_0$ partial wave we fix the value of
the scattering length~\footnote{
Of course, using the term scattering {\it length} for a $p$-wave
is language abuse.
The proper name is scattering volume, as it has dimensions of
${[ \rm length ]}^3$.
A similar comment would apply in the d-wave case, in which the so-called
scattering {\it length} has dimension of ${[ \rm length ]}^5$ and
henceforth is a (5-dimensional) hypervolume.
However, for notational simplicity we will always use the term scattering
length for the coefficient related to the low energy behaviour of
the $l$-wave phase shift,
$\delta_l(k) = - a_l\,k^{2 l+ 1} + \mathcal{O}(k^{2 l+3})$.
},
see the appendix \ref{app:LO} for the details.
Of course, there is still a residual cut-off dependence
that is however mostly harmless owing to the existence
of a clear convergence pattern in the $r_c \to 0$ limit.
That is, we can interpret the residual cut-off dependence
as a higher order effect.
For the regularization employed in this case, 
the cut-off dependence of the phase shifts can be explicitly
computed by employing the methods of Ref.~\cite{PavonValderrama:2007nu},
yielding the result
\begin{eqnarray}
\frac{d\,\delta^{(0)}_{{}^3P_0}(k; r_c)}{d r_c} =
k^3\,{\left| \frac{u_k^{(0)}(r_c)}{k} \right|}^2\,
\sim k^3 \, r_c^{3/2} \, a_3^{1/2} \, ,
\label{eq:3P0-rc-dependence}
\end{eqnarray}
which serves as a formal confirmation of the existence of
the $r_c \to 0$ limit.

\subsection{Subleading Order Modifications of the Counting}
\label{sub:n2lo}

\begin{figure*}[htt!]
\begin{center}
\epsfig{figure=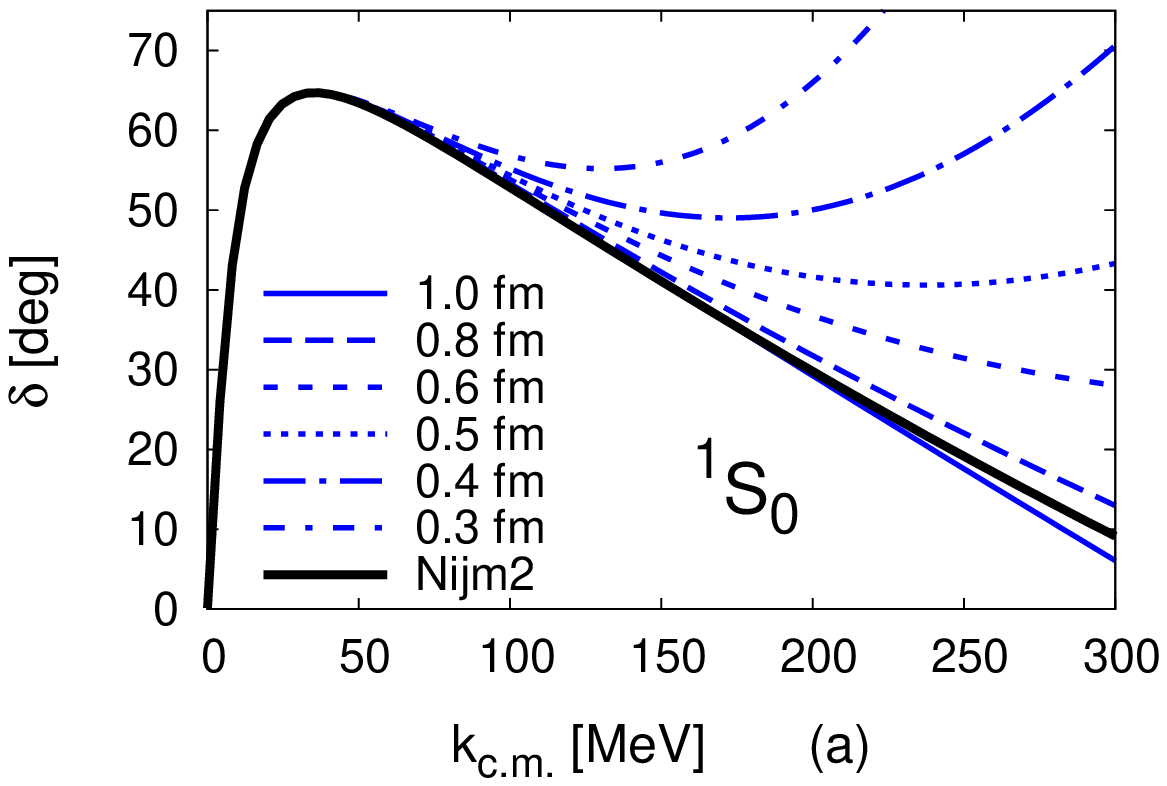,
	height=4.75cm, width=8.0cm}
\epsfig{figure=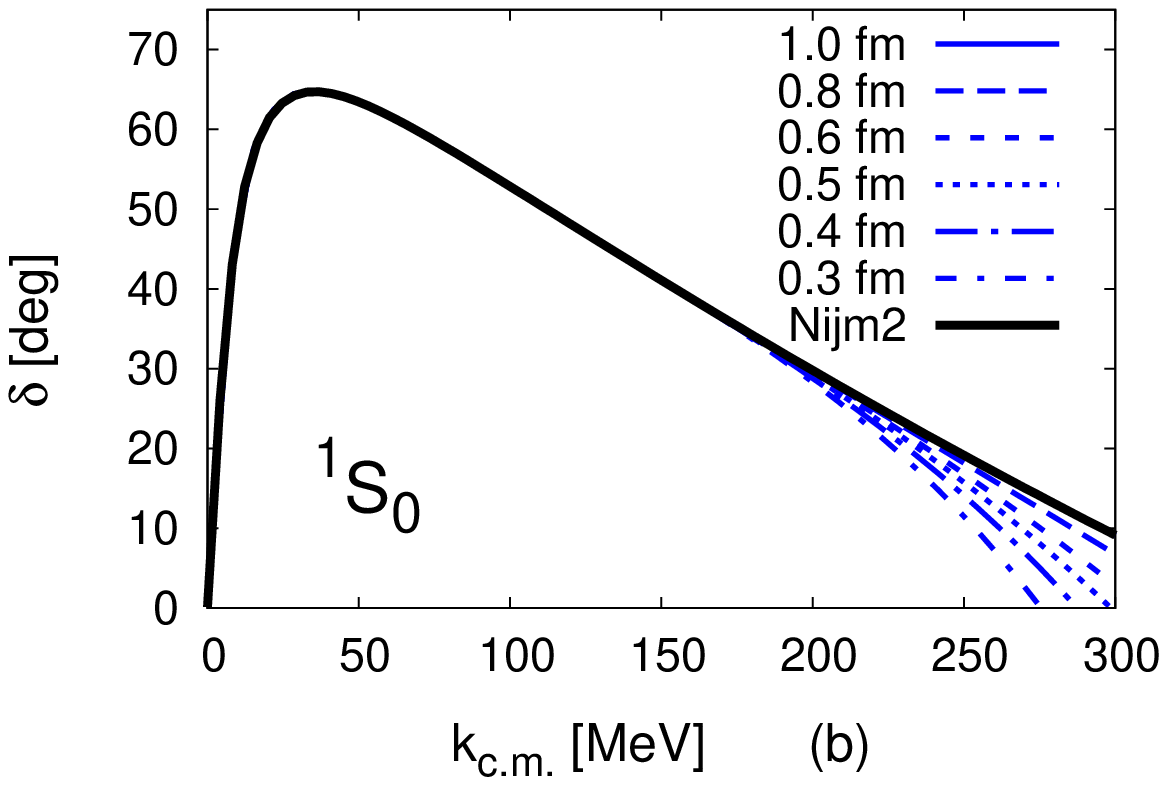, 
	height=4.75cm, width=8.0cm}
\end{center}
\caption{(Color online)
Next-to-next-to leading order phase shifts for the $^1S_0$ channel
in the perturbative Weinberg counting (left panel) and in the
Nogga, Timmermans and van Kolck scheme~\cite{Nogga:2005hy}
(right panel).
By perturbative Weinberg we mean that the subleading pieces of the potential
(that is, chiral two pion exchange) are treated as perturbations.
We can appreciated that the cut-off dependence appearing in the perturbative
Weinberg set-up can be cured by the inclusion of an additional counterterm.
}
\label{fig:modified-weinberg-subleading}
\end{figure*}

The modifications of the power counting induced by renormalizability
are not confined to leading order.
Further deviations occur at subleading orders, which can be identified
by analyzing the cut-off dependence of the phase shifts,
even though these orders are included as perturbations.
In this case, the most direct example is the $^1S_0$ channel,
which follows the Weinberg counting at ${\rm LO}$,
but diverts from it when subleading
corrections are included.
At orders $Q^2$ and $Q^3$ the Weinberg counting dictates a total of
two counterterms for the $^1S_0$ channel, which in momentum space
take the form
\begin{eqnarray}
\label{eq:contact-W}
\langle p | V^{(\nu = 2,3)}_C | p' \rangle = C_0^{(\nu)} + C_2^{(\nu)} (p^2 + p'^2)
\, .
\end{eqnarray}
In terms of renormalization, this counterterm structure is equivalent to
fixing two observables, for example the scattering length and
the effective range of the $^1S_0$ channel.
For making such a calculation we will follow the formalism of
Ref.~\cite{Valderrama:2009ei}, but adapting the number of
counterterms to two.
In such a case, we obtain the phase shifts of
Fig.~(\ref{fig:modified-weinberg-subleading}).

As can be appreciated, there is a strong cut-off dependence
in the results of Fig.~(\ref{fig:modified-weinberg-subleading}).
The reason can be found by analyzing the cut-off behaviour of the
perturbative corrections to the phase shifts.
According to Ref.~\cite{Valderrama:2009ei},
the $Q^{\nu}$ contribution to the phase shift is proportional to the integral
\begin{eqnarray}
\label{eq:pert-1S0}
\delta^{(\nu)}(k; r_c) \propto \frac{\mu}{k}
\int_{r_c}^{\infty}\,dr\, V^{(\nu)}(r) {u_k^{(0)}\,}^2(r) \, ,
\end{eqnarray}
where $u_k^{(0)}(r)$ is the ${\rm LO}$ wave function in the Weinberg counting,
which can be described by adapting the Schr\"odinger equation
for the $^3P_0$ channel, Eq.~(\ref{eq:bc-3P0}),
to the $^1S_0$ case,
and $V^{(\nu)}$ is the order $Q^{\nu}$ contribution to the chiral potential.
It should be noted that $\delta^{(\nu)}$ scales as $Q^{\nu + 1}$,
even though nominally it is of order $Q^{\nu}$.
The degree of divergence of $\delta^{(\nu)}$ can be easily determined
by considering the Taylor expansion of the ${\rm LO}$ wave function
~\cite{PavonValderrama:2005wv}, that is
\begin{eqnarray}
u_k^{(0)}(r) = u_0 + k^2 u_2 + k^4 u_4 + \dots \, ,
\end{eqnarray}
where the behaviour of the different terms at short enough distances
is given by  $u_0 \sim 1$, $u_2 \sim r^2$, $u_4 \sim r^4$, and so on.
With the previous information, and the additional fact
that $V^{(\nu)} \sim 1/r^{3+\nu}$,
we find that for the particular case $\nu = 3$
the integral in Eq.~(\ref{eq:pert-1S0}) diverges as
\begin{eqnarray}
I^{(\nu = 3)}(k; r_c) &=&
\int_{r_c}^{\infty} \,dr\,V^{(\nu)}(r) {u_k^{(0)}\,}^2(r) \nonumber \\
&=&
{\left( \frac{a}{r_c} \right)}^5 \,c_0 + 
(k a)^2 {\left( \frac{a}{r_c} \right)}^3 \,c_2 \nonumber \\ &+& 
(k a)^4 {\left( \frac{a}{r_c} \right)} \,c_4
+ \dots \, ,
\end{eqnarray}
with $a$ the length scale governing the divergences and  
where $c_0$, $c_2$ and $c_4$ are dimensionless numbers
that are expected to be of order one,
while the terms in the dots are already finite.
In Fig.~(\ref{fig:modified-weinberg-subleading})
we have only fixed the $k^0$ and $k^2$
behaviour for reproducing the low energy phase shifts, that is,
the scattering length and the effective range.
This means that there is still a contribution that diverges as $k^4/r_c$ 
explaining the failure of the Weinberg counting
in the perturbative context.

Again, the solution is to modify the counting
in such a way as to include a new counterterm.
This means that the correct $Q^2$ and $Q^3$ contact interaction
should be~\footnote{We have ignored the ${C_4'}^{(\nu)} p^2 p'^2$
operator, as it is redundant~\cite{Beane:2000fi}.
However, the equivalence of this operator with $C_4 (p^4 + p'^4)$
only shows up either in dimensional regularization or
once the floating cut-off is removed.
For finite cut-off calculations, the difference between the $C_4$ and $C_4'$
operators is presumably a higher order effect.
On the practical side the inclusion of this operator will surely
be advantageous in momentum space treatments as it may help to
reduce the cut-off dependence of the scattering observables.
}
\begin{eqnarray}
\label{eq:contact-NTvK}
\langle p | V^{(\nu = 2,3)}_C | p' \rangle &=&
C_0^{(\nu)} + C_2^{(\nu)} (p^2 + p'^2) \nonumber \\
&+& C_4^{(\nu)} (p^4 + p'^4) \, ,
\end{eqnarray}
instead of the form dictated by Weinberg dimensional counting in
Eq.~(\ref{eq:contact-W}).
The additional counterterm is able to absorb the $k^4/r_c$ 
divergence in the ${\rm N^2LO}$ phase shift.
In such a case, the new cut-off dependence is the one given by
Fig.~(\ref{fig:modified-weinberg-subleading}),
in which there is still a residual cut-off dependence
of ${\mathcal O}(k^6 r_c)$.
However, the same comments apply as in the $^3P_0$ channel at ${\rm LO}\,$:
the residual cut-off dependence is a higher order effect
and can be safely ignored.

Of course, the power counting modification of Eq.~(\ref{eq:contact-NTvK})
is well-known to arise in two-body systems with a large value
of the scattering length~\cite{Kaplan:1998tg,Kaplan:1998we,Birse:1998dk,Barford:2002je},
where the renormalization group analysis
of the $C_4$ operator indicates that its actual
scaling is $Q^2$ rather than the naive dimensional expectation $Q^4$.
In this regard, it should be stressed that
the $^1S_0$ phase shifts cannot be properly
renormalized at ${\rm NLO}$/${\rm N^2LO}$
without the inclusion of the $C_4$ operator.
The previous is not merely a statement about the existence of the
$r_c \to 0$ limit in the calculation,
but also about the accuracy of the theoretical description of the system.
The renormalized calculation is accurate up to
$Q^3$($Q^4$) at ${\rm NLO}$ (${\rm N^2LO}$).
On the contrary, if we do not to include the $C_4$ operator, the error
of the calculation will be $\mathcal{O} (Q^2)$,
much bigger than expected.
This may prove to be a significant drawback for 
the nuclear EFT lattice approach~\cite{Borasoy:2007vi,Borasoy:2007vk,Epelbaum:2008vj,Epelbaum:2009zsa,Epelbaum:2009pd,Epelbaum:2010xt,Epelbaum:2011md},
which does not include the aforementioned operator
in the $^1S_0$ channel.

Finally,
the observation that the contact interaction of Eq.~(\ref{eq:contact-NTvK})
renormalizes the scattering amplitude can be employed to extract
the breakdown scale of the EFT in the singlet channel,
as has been shown by the deconstruction
approach of Refs.~\cite{Birse:2007sx,Birse:2010jr}.
The idea is to employ the phenomenological phase shifts as the input
of a calculation for determining the form of the contact range
interaction up to an error of ${\mathcal O}(Q^{\nu + 1})$.
This represents in fact the complementary process of fitting the
$C_{2n}^{(\nu)}$ parameters to the phase shifts.
By considering the on-shell form of the contact range potential,
$V_C^{(\nu)}(k) = \langle k | V^{(\nu)}_C | k \rangle$,
Ref.~\cite{Birse:2010jr} is able to determine that (i) the form of the
short range interaction complies with Eq.~(\ref{eq:contact-NTvK})
and (ii) the hard scale governing $V_C$ is
$\Lambda_{0,s} \simeq 270\,{\rm MeV}$.
This figure is much lower than the expected $\Lambda_{0} \sim 0.5\,{\rm GeV}$
and translates into a slow convergence rate for the singlet channel,
$Q / \Lambda_{0} \simeq m_{\pi} / \Lambda_{0,s} \simeq 0.5$,
emphasizing the importance of the proper renormalization of
the scattering amplitude to minimize the theoretical uncertainty.
The breakdown scale can in fact be appreciated in the right panel
of Fig.~(\ref{fig:modified-weinberg-subleading}),
where the cut-off dependence of the renormalized results starts to become
sizable at $k \simeq \Lambda_{0,s}$, in agreement with the estimates
of Ref.~\cite{Birse:2010jr}~\footnote{A similar comment applies for the phase
shifts computed in the Weinberg counting: if we inspect the left panels of
Figs.~(\ref{fig:modified-weinberg-leading}) and (\ref{fig:modified-weinberg-subleading}),
the cut-off dependence is already conspicuous at $k \simeq 100-200\,{\rm MeV}$,
suggesting a relatively soft breakdown scale for both the $^1S_0$ and $^3P_0$
partial waves.
In contrast, 
for the $^3P_0$ channel in the Nogga et al. counting~\cite{Nogga:2005hy}
the cut-off dependence of the right panel of
Fig.~(\ref{fig:modified-weinberg-leading})
does not provide a clear indication of the breakdown scale.
In this case the determination of the range of applicability of the theory
is better done by analyzing the size the subleading order corrections
to the phase shift.
}.

\section{Perturbative Renormalizability of Singular Potentials
and Power Counting}
\label{sec:pert-reno}

In this section we study the power counting that arises when one pion exchange
is iterated to all orders and chiral two-pion exchange is treated
as a perturbation.
We follow a more formal and detailed approach than in the previous section.
We start by defining the power counting conventions
in Sect. \ref{subsec:conventions}.
We derive the power counting by requiring the perturbative corrections
to remain finite in the $r_c \to 0$ limit,
first for the uncoupled channels in Sect. \ref{subsec:uncoupled}
and then we extend the results to the coupled channels
in Sect. \ref{subsec:coupled}.
We informally comment on the convergence rate and the expansion parameter
of the resulting EFT in Sect. \ref{subsec:convergence}.
Finally, we discuss the perturbative power counting
in Sect. \ref{subsec:counting}.
The results of this section are briefly summarized
in Table \ref{tab:counterterms},
where we can check the number of counterterms required
in the lower partial waves from ${\rm LO}$ up to ${\rm N^3LO}$.

\subsection{Conventions}
\label{subsec:conventions}

We assume that the chiral potential is expanded in terms of a power counting
as follows
\begin{eqnarray}
\label{eq:pot-expansion}
V(\vec{q}) = \sum^{\nu_{\rm max}}_{\nu = \nu_0} V^{(\nu)}(\vec{q}) +
\mathcal{O}\left( (\frac{Q}{\Lambda_0})^{\nu_{\rm max} + 1}\right)\, ,
\end{eqnarray}
where $Q$ ($\Lambda_0$) is the soft (hard) scale of the system.
The expansion starts at order $\nu_0 \geq -1$, and we only consider
the corrections to the chiral potential up to a certain order
$\nu_{\rm max}$ ($=2,3$ in the present paper).
The T-matrix follows the same power counting expansion
as the potential, that is
\begin{eqnarray}
\label{eq:amp-expansion}
T = \sum^{\nu_{\rm max}}_{\nu = \nu_0} T^{(\nu)} +
\mathcal{O}\left( (\frac{Q}{\Lambda_0})^{\nu_{\rm max} + 1}\right)\, .
\end{eqnarray}
The scattering equations can be obtained by reexpanding
the Lippmann-Schwinger equation
\begin{eqnarray}
T = V + V G_0 T \, ,
\end{eqnarray}
in terms of the power counting of the potential, taking into account
that the resolvent operator $G_0$ is formally of order $Q$.
In this convention, only contributions to the chiral potential
which are of order $Q^{-1}$ should be iterated,
yielding the equation
\begin{eqnarray}
T^{(-1)} &=& V^{(-1)} + V^{(-1)} G_0 T^{(-1)} \, .
\end{eqnarray}
The corresponding dynamical equation of the higher order contributions
to the T-matrix are increasingly more involved.

In principle, in the Weinberg scheme the power counting expansion
starts at order $Q^0$
\begin{eqnarray}
V_{\rm NN} = V^{(0)} + V^{(2)} + V^{(3)} + \mathcal{O}(Q^4) \, .
\end{eqnarray}
Within the power counting conventions we follow, the previous means
that all pieces of the potential should be treated as perturbations.
However, it is clear that $V^{(0)}$ needs to be iterated
in certain partial waves.
This requires a promotion from order $Q^0$ to $Q^{-1}$, that is
\begin{eqnarray}
V^{(0)} \to V^{(-1)} \, ,
\end{eqnarray}
at least in the partial waves in which either the ${\rm LO}$ contact
operator or OPE are non-perturbative.
From a purely quantum-mechanical point of view the iteration
is necessary if the potential is not weak.
In terms of power counting, a large coupling constant can be accounted for
by identifying it with $\Lambda_0 / Q$, which is a large number.
Alternatively, we can try to explicitly identify the low energy scale which
requires the ${\rm LO}$ potential to be iterated.
For the ${\rm LO}$ contact operator
the additional low energy scale is well known to be
the inverse of the large scattering length of
the $^1S_0$ and $^3S_1-{}^3D_1$ channels,
as has been extensively discussed
in the literature~\cite{Kaplan:1998tg,Kaplan:1998we,Gegelia:1998gn,Birse:1998dk,vanKolck:1998bw}.
For justifying the iteration of the OPE potential
Birse has proposed~\cite{Birse:2005um}
the identification of
\begin{eqnarray}
\Lambda_{\rm OPE} = \Lambda_{\rm NN} = \frac{16 \pi f_{\pi}^2}{M_N g_A^2}
\simeq 300\,{\rm MeV}
\end{eqnarray}
as the new low energy scale.
However, this choice does not explain the fact that OPE remains perturbative
in the singlet channels even if iterated~\cite{Fleming:1999ee}.
A different possibility may be to treat the inverse
of the length scale governing the strength of the tensor force
as the missing low energy scale, giving a $\Lambda_{\rm OPE}$
that is in general a fraction of $\Lambda_{\rm NN}$.

After the promotion of ${\rm LO}$ potential from order $Q^0$ to order $Q^{-1}$,
we encounter a remarkable simplification in the dynamical equations
describing the $Q^2$ and $Q^3$ contributions to the T-matrix,
namely
\begin{eqnarray}
T^{(\nu)} &=& (1 + T^{(-1)} G_0 ) V^{(\nu)} (G_0 T^{(-1)} + 1) \, , 
\end{eqnarray}
where, due to the absence of contributions to the finite range chiral potential
from order $Q^{-1}$ to order $Q^{2}$, second order perturbation
theory is not needed up to order $Q^5$.
That is, the equation above is valid for $\nu < 5$.

\begin{table}
\begin{center}
\begin{tabular}{|c|c|c|c|c|}
\hline \hline
Partial wave & ${\rm LO}$ & ${\rm NLO}$ & ${\rm N^2LO}$ & ${\rm N^3LO}$ \\ 
\hline
$^1S_0$ & $1$ & $3$ & $3$ & $4$ \\
$^3S_1-{}^3D_1$ & $1$ & $6$ & $6$ & $6$ \\ \hline
$^1P_1$ & $0$ & $1$ & $1$ & $2$ \\
$^3P_0$ & $1$ & $2$ & $2$ & $2$ \\
$^3P_1$ & $0$ & $1$ & $1$ & $2$ \\
$^3P_2-{}^3F_2$ & $1$ & $6$ & $6$ & $6$ \\ \hline
$^1D_2$ & $0$ & $0$ & $0$ & $1$ \\
$^3D_2$ & $1$ & $2$ & $2$ & $2$ \\
$^3D_3-{}^3G_3$ & $0$ & $0$ & $0$ & $1$ \\ \hline
All & 5 & 21 & 21 & 27 \\
\hline \hline
\end{tabular}
\end{center}
\caption{
Total number of counterterms per partial wave at ${\rm LO}$ ($Q^{-1}$),
${\rm NLO}$ ($Q^2$), ${\rm N^2LO}$ ($Q^3$) and ${\rm N^3LO}$ ($Q^4$)
assuming (i) the iteration of the OPE potential at ${\rm LO}$ in
the $^3S_1-{}^3D_1$, $^3P_0$, $^3P_2-{}^3F_2$ and $^3D_2$ partial
waves and (ii) energy-dependent counterterms.
The ${\rm N^3LO}$ results assume that an energy-independent representation
of the finite-range chiral potential is being used.
If an energy-dependent potential is employed, the number of free parameters
will grow from $27$ to $35$ (see the discussion in Sect.\ref{subsec:counting}
for details).
} \label{tab:counterterms}
\end{table}

\subsection{Uncoupled Channels}
\label{subsec:uncoupled}

We consider in the first place the power counting expansion of
the $l$-wave phase shift
\begin{eqnarray}
\delta_l(k; r_c) = \sum^{\nu_{\rm max}}_{\nu = -1} \delta_l^{(\nu)}(k; r_c) 
+ {\mathcal O}(Q^{\nu_{\rm max} + 1}) \, .
\end{eqnarray}
The ${\rm LO}$ phase shift is calculated non-perturbatively by solving
the reduced Schr\"odinger equation with the ${\rm LO}$ chiral potential,
see Appendix~\ref{app:LO} for details.
On the other hand, we compute the order $\nu$-th contribution
to the phase shift, $\delta_l^{(\nu)}$,
in the distorted wave Born approximation (DWBA), that is
\begin{eqnarray}
\frac{\delta_l^{(\nu)}(k; r_c)}{\sin^2{\delta_l^{(-1)}}} 
= - \frac{2\mu}{k^{2 l + 1}}\,{\mathcal{A}_l^{(-1)}}^2\,I^{(\nu)}_l(k; r_c)
\end{eqnarray}
where $\delta_l^{(-1)}$ is the leading order phase shift, $\mu$ the reduced
mass of the two nucleon system, $k$ the center-of-mass momentum, $l$ the
angular momentum of the system, ${\mathcal{A}_l^{(-1)}}$ a normalization
factor and $I^{(\nu)}_l$ the perturbative integral.
The perturbative integral $I^{(\nu)}_l$ is defined as
\begin{eqnarray}
I^{(\nu)}_l(k; r_c) = \int_{r_c}^{\infty}\,dr\,V^{(\nu)}(r)\,{u_{k,l}^{(-1)}(r)\,}^2
\end{eqnarray}
where $u_{k,l}^{(-1)}$ is the ${\rm LO}$ wave function, which we 
assume to be asymptotically ($r \to \infty$) normalized to
\begin{eqnarray}
\frac{\mathcal{A}_l^{(-1)}}{k^l}\,u^{(-1)}_{k,l}(r) \to
\cot{\delta_l^{(-1)}}\,\hat{j}_l(k r) - \hat{y}_l (k r) \, ,
\end{eqnarray}
where $\hat{j}_l(x) = x j_l(x)$, $\hat{y}_l(x) = x y_l(x)$,
with $j_l(x)$ and $y_l(x)$ the spherical Bessel functions.
For the purpose of analyzing the renormalization of the perturbative
phase shifts, we will assume that the normalization factor
${\mathcal{A}_l^{(-1)}}(k)$ is defined in such a way that
the ${\rm LO}$ wave function is energy-independent
at $r \to 0$, that is
\begin{eqnarray}
\label{eq:uk-norm-short}
\lim_{r \to 0}\frac{d^2}{dk^2}\,u_{k,l}(r)  = 0 \, ,
\end{eqnarray}
a prescription that will make relatively easy the identification of the
necessary number of counterterms.
However, in practical calculations we will simply take
\begin{eqnarray}
\label{eq:uk-norm-short-prac}
\frac{d^2}{dk^2}\,u^{(-1)}_{k,l}(r) \Big|_{r=r_c}  = 0 \, ,
\end{eqnarray}
which is easier to implement than Eq.~(\ref{eq:uk-norm-short}).

The renormalizability of the perturbative phase shifts can be easily analyzed
in terms of the short range behaviour of the ${\rm LO}$ reduced wave function 
and the $Q^{\nu}$ contribution to the chiral potential,
which determine whether the perturbative integral
$I_l^{(\nu)}$ is divergent or not.
In particular, for $m r \ll 1$, we have the following
\begin{eqnarray}
u^{(-1)}_{k,l}(r) \sim r^{s} \quad,\quad V^{(\nu)}(r) \sim \frac{1}{r^{3+\nu}}
\, ,
\end{eqnarray}
where $s$ describes the power-law behaviour of the wave function at
short distances.
For the perturbative integral we have
\begin{eqnarray}
I_l^{(\nu)}(k,r_c) \sim \int_{r_c} \frac{d\,r}{r^{3+\nu-2s}} \, ,
\end{eqnarray}
which diverges if $3 + \nu - 2s \geq 1$.
The necessary number of counterterms can be determined by considering
the Taylor expansion of the perturbative integral in terms of $k^2$
\begin{eqnarray}
I_l^{(\nu)}(k,r_c) = \sum_{n=0}^{\infty}\,I_{2n,l}^{(\nu)}(r_c) k^{2n} \, ,
\end{eqnarray}
in which each new term is less singular than the previous one
as a consequence of the related expansion for
the ${\rm LO}$ reduced wave function
\begin{eqnarray}
\label{eq:uk_exp}
u^{(-1)}_{k,l}(r) = \sum_{n=0}^{\infty}\,u^{(-1)}_{2n, l}(r) k^{2n} \, ,
\end{eqnarray}
where $u^{(-1)}_{2n, l} \sim r^{s + n t}$ for small radii,
with $t \geq 2$.
We can see now that the reason for choosing the energy-independent normalization
of Eq.~(\ref{eq:uk-norm-short}) is to have additional power law
suppression at short distances in the $k^2$ expansion of
the wave function.
In terms of the expansion of the perturbative integral, we obtain
\begin{eqnarray}
I_{2n,l}^{(\nu)}(r_c) \sim \int_{r_c} \frac{d\,r}{r^{3+\nu-2s-n\,t}} \, ,
\end{eqnarray}
which converges for $3 + \nu - 2s - n\,t < 1$.
Eventually, as $n$ increases, we will have well-defined integrals.
In this regard, we can define the number of subtraction/counterterms
as the smallest value of $n_c$ for which
\begin{eqnarray}
3 + \nu - 2s - n_c\,t < 1 \, .
\end{eqnarray}
It should be noted that $n_c$ refers to the total number of
subtraction/counterterms required at a given order.
For example, if $n_c = 1$ the counterterm only affects the scattering
length of the $\nu$-th order of the phase shift.
If the scattering length was already fixed at a lower order, this just
indicates that a perturbative correction must be added to the lower
order counterterm.

To summarize, we have that the perturbative integral can be divided
into a divergent and a regular piece
\begin{eqnarray}
I^{(\nu)}_l(k; r_c) = I^{(\nu)}_{l,D}(k; r_c) + I^{(\nu)}_{l,R}(k; r_c) \, ,
\end{eqnarray}
where the divergent piece can be expanded in powers of the squared
momentum $k^2$,
\begin{eqnarray}
I^{(\nu)}_{l,D}(k; r_c) = \sum_{n = 0}^{n_c-1} I^{(\nu)}_{l,2n}(r_c)\,k^{2n}
\, ,
\end{eqnarray}
with $n_c$ the number of counterterms that renders
the perturbative phase shifts finite.
The perturbative integral can be regularized in any specific way
which we find convenient.
A particularly simple and straightforward method is to add
$n_c$ free parameters to the original integral as follows
\begin{eqnarray}
\label{eq:I_nu_modified}
\hat{I}^{(\nu)}_l(k; r_c) = \sum_{n = 0}^{n_c-1} \lambda^{(\nu)}_{l,2n}\,k^{2n}
 + I^{(\nu)}_{l}(k; r_c) \, ,
\end{eqnarray}
where the parameters $\lambda^{(\nu)}_{l,2n}$ are to be fitted to
the experimental phase shifts.
The previous parameters, which we will informally call $\lambda$-parameters,
can be related to the more standard counterterms
if we consider a concrete representation of the short range physics,
for example
\begin{eqnarray}
V_{C,l}^{(\nu)}(r; r_c) = \frac{f_l^2(r_c)}{4 \pi r_c^2}\,
\sum_{n = 0}^{n_c-1}\,C^{(\nu)}_{2n,l}(r_c) k^{2n}\,\delta(r-r_c) \, , 
\nonumber \\
\end{eqnarray}
where $f_l(r_c) = \frac{(2l+1)!!}{r_c^l}$ is a factor which ensures
that the matrix elements of the previous potential in momentum
space behaves as
\begin{eqnarray}
\langle p | V_{C,l} | p' \rangle \to p^l {p'}^l\,
\sum_{n}\,C^{(\nu)}_{2n,l}(r_c) k^{2n} \, ,
\end{eqnarray}
for $r_c \to 0$.
If we have chosen to work in the practical normalization of
Eq.~(\ref{eq:uk-norm-short-prac}),
the following relationship is obtained between the fitting parameters
$\lambda^{(\nu)}_{2n,l}$ and the usual counterterms
\begin{eqnarray}
\lambda^{(\nu)}_{2n,l} = \frac{f_l^2(r_c)}{4 \pi r_c}\,C^{(\nu)}_{2n,l}(r_c)\,
u^{(-1)}_{0,l}(r_c) \, .
\end{eqnarray}
For other representations of the short range physics, the relationship
will take a more complex form.

We can distinguish four cases: (i) the irregular solution of a non-singular
potential ($^1S_0$), (ii) the regular solution of a non-singular potential
($^1P_1$, $^1D_2$), (iii) the general solution of an attractive singular
potential ($^3P_0$, $^3D_2$) and (iv) the regular solution
of a repulsive singular potential ($^3P_1$).
In the following paragraphs we will discuss each of these cases in detail.

\subsubsection{Non-Singular Potential}

Non-singular (or regular) potentials are potentials for which
the regularity condition is fulfilled
\begin{eqnarray}
\lim_{r \to 0} r^2 V(r) = 0 \, .
\end{eqnarray}
The regularity condition implies in turn that the short range behaviour
of the wave function is determined by the centrifugal barrier,
which overcomes the potential at short distances.
For $r \to 0$, we can define a regular and irregular solutions
which behave as
\begin{eqnarray}
u^{(-1)}_{l,k,{\rm reg}}(r) &\sim& r^{l + 1} \, , \\
u^{(-1)}_{l,k,{\rm irr}}(r) &\sim& \frac{1}{r^{l }} \, .
\end{eqnarray}
If there is no short range physics at ${\rm LO}$,
the regular solution is chosen.
On the contrary, if a contact operator is included at ${\rm LO}$,
the ${\rm LO}$ reduced wave function will be a superposition
of the regular and irregular solutions.

The regular solution gives rise to the Weinberg power counting unaltered.
If we consider the $k^2$ expansion of the regular solution,
Eq.~(\ref{eq:uk_exp}), we have
\begin{eqnarray}
u^{(-1)}_{l,k,{\rm reg}} \sim r^{l + 2 n + 1} \, ,
\end{eqnarray}
that is, $s = l + 1$ and $t = 2$.
The convergence of the perturbative integral requires in this case
\begin{eqnarray}
\nu < 2 l + 2 n_c \, ,
\end{eqnarray}
where $n_c$ stands for the number of counterterms
in a given partial wave.
Therefore new counterterms appear as expected in naive dimensional analysis,
that is
\begin{eqnarray}
C_{2n,l} \sim Q^{2l+2n} \, .
\end{eqnarray}
For the particular case of the $^1P_1$ channel, the first counterterm
($C_{0,{{}^1P_1}}$) is required at $Q^2$, a second ($C_{2,{{}^1P_1}}$)
at $Q^4$, and so on.

In contrast, the occurrence of the irregular solution at ${\rm LO}$ will
change the power counting.
The $k^2$ expansion of the wave function yields in this case
\begin{eqnarray}
u^{(-1)}_{l,k,{\rm irr}} \sim r^{-l + 2 n} \, ,
\end{eqnarray}
which implies $s = -l$ and $t = 2$.
The finiteness condition reads
\begin{eqnarray}
2 n_c > \nu + 2 l + 2 \, ,
\end{eqnarray}
which implies a significant deviation from naive dimensional analysis.
In fact the scaling of the counterterms is given by
\begin{eqnarray}
C_{2n,l} \sim Q^{2n - 2l - 2} \, .
\end{eqnarray}
For the $^1S_0$ channel, the previous requires that the first perturbative
correction to the $C_{0,{^1S_0}}$ counterterm enters at order $Q^{-2}$,
while the $C_{2n,{^1S_0}}$ counterterms (with $n > 0$) are
required at order $Q^{2n - 2}$.
It should be noticed that the order $Q^{-2}$ assignment to the first
perturbative correction to the $C_0$ counterterm is merely formal:
this correction actually enters at the same order as the first
perturbative correction to the finite range piece of
the potential, which happens at order $Q^2$.
In fact, the $Q^{-2}$ scaling is rather an indication of the fine-tuning
required for having an unnatural scattering length.
In terms of Birse's renormalization group analysis (RGA)~\cite{Birse:1998dk},
the $Q^{-2}$ eigenvalue is a reflection of the unstable
nature of the infrared fixed point associated with
two-body systems with a large scattering length.

It is interesting to notice that the extension of the power counting
for irregular solutions to $p$-waves leads to the same conclusions
as p-wave Halo EFT~\cite{Bertulani:2002sz}, namely,
that the first two contact operators should be iterated.
The previous arguments imply that for a $p$-wave, the first counterterm
will be of order $Q^{-4}$, the second $Q^{-2}$, and the third $Q^0$.
This has two interpretations: first, that the unstable infrared fixed point
is even more fine-tuned than in $s$-waves, and second, that it may be
necessary to iterate two counterterms instead of only one.
However, for the particular case of nucleon-nucleon scattering,
we do not encounter this situation.

\subsubsection{Singular Potential}

The appearance of power-law singular interactions at ${\rm LO}$
entails substantial changes to the power counting of
the short range operators.
In particular we are interested in the effect of tensor OPE, 
which at short distances ($r \to 0$) behaves as
\begin{eqnarray}
2\mu\,V^{(-1)}_{\rm tensor}(r) \to \pm \frac{a_3}{r^3} \, , 
\end{eqnarray}
where $\mu$ is the reduced mass of
the two-nucleon system and $a_3$ a length scale that governs
the strength of the tensor force.
We have chosen a convention in which $a_3 > 0$ and where the attractive or
repulsive character of the potential is indicated by
the explicit $\pm$ sign.
The renormalization of singular potentials implies that attractive
singular interactions require the inclusion of a counterterm,
while repulsive singular interactions do not~\cite{PavonValderrama:2005gu,PavonValderrama:2005wv,PavonValderrama:2005uj},
a feature which will be reviewed in the following paragraph.

For a tensor force, the terms of the $k^2$ expansion of the wave function 
behave for $r \to 0$ as
\begin{eqnarray}
u^{(-1)}_{l,2n}(r) \simeq r^{3/4 + 5 n / 2} f(2 \sqrt{\frac{a_3}{r}}) \, .
\end{eqnarray}
If the tensor force is attractive, the function $f$ takes the form 
\begin{eqnarray}
f_A(x) \sim C_S \sin{x} + C_C \cos{x} \, ,
\end{eqnarray}
with $x = 2 \sqrt{\frac{a_3}{r}}$ and
where the specific linear combination of sine and cosine factors is
determined by the ${\rm LO}$ counterterm.
In contrast, for a repulsive tensor force we have a regular
and irregular solution.
We have indeed that
\begin{eqnarray}
f_R(x) \sim C_{\rm reg} e^{-x} + C_{\rm irr} e^{+x} \, .
\end{eqnarray}
where the $e^{-x}$ solution is regular at the origin,
while the $e^{+x}$ solution diverges exponentially for $r \to 0$.
For the repulsive case we will always chose the regular solution at ${\rm LO}$.

In the attractive case, the convergence of the perturbative integrals is
solely determined by the power-law behaviour of the reduced wave
functions: the sine and cosine factors do not play any role
on the finiteness of the perturbative corrections.
Renormalizability requires 
\begin{eqnarray}
\frac{5}{2}\, n_c > \nu + \frac{1}{2} \, ,
\end{eqnarray}
which in turn implies that the first perturbative correction
to the $C_{2n,l}$ counterterm scales as
\begin{eqnarray}
C_{2n,l} \sim Q^{(5 n- 1) / 2} \, ,
\end{eqnarray}
that is, the first correction to $C_0$ enters at order $Q^{-1/2}$, 
the $C_2$ counterterm at order $Q^2$, and so on.
It is interesting to notice that (i) the power counting of the counterterms
does not depend on the partial wave considered, (ii) the first perturbative
correction to $C_0$ is ${\mathcal O}(Q^{-1/2})$, which means that the
infrared fixed point is stable, and (iii) the appearance of new
counterterms is delayed in comparison with the non-singular potentials.

In contrast, the analysis for repulsive tensor force is puzzling.
The decreasing exponential behaviour of the wave function at short distances
is able to render the perturbative integral finite,
independently of the degree of singularity of 
he $\nu$-th order potential.
This feature can be easily appreciated by considering the integral
\begin{eqnarray}
I^{(\nu)} \sim 
\int_{r_c} \frac{dr}{r^{3+\nu - 3/2}}\,\exp{(-4 \sqrt{\frac{a_3}{r}})} \, ,
\end{eqnarray}
which is always finite.
However, for a consistent power counting to emerge
we need new counterterms at higher orders.
Two solutions are possible: (i) follow the power counting rules for
attractive singular interaction, that is, $C_{2n} \sim Q^{(5 n-1) / 2}$,
(ii) follow naive dimensional analysis, i.e. Weinberg counting,
leading to $C_{2n,l} \sim Q^{2n+2l}$.

The first solution was proposed by Birse in Ref.~\cite{Birse:2005um}.
It only considers as relevant to the power counting the power-law behaviour
of the wave function, but not the sine, cosine or exponential factor.
This expectation is consistent with the fact that the exponentials
are just a sine or cosine function with an imaginary argument.
However, it is also natural to expect that short distances do not play
a significant role if the particles cannot approach each other
due to the repulsive potential barrier.

This observation leads to the second solution, namely
to follow naive dimensional analysis, which implicitly assumes
that the repulsive channel can be treated as a perturbation~\footnote{
Alternatively, we can try to analyze the scaling of
the wave function at distances of the order of the breakdown
scale of the theory, $r \simeq 0.5\,{\rm fm}$.
If the wave function behaves as $r^{l+1}$ (instead of
the singular $r^{3/4}\,{\rm exp}(-2\sqrt{a_3/r})$),
naive dimensional expectations are fulfilled and
the use of Weinberg counting is justified.
Unfortunately, for the particular case of the $^3P_1$ channel
the singular behaviour is achieved at distances
of about $1\,{\rm fm}$ (as $a_3 = 1.34\,{\rm fm}$ in this case),
so the wave function scaling argument cannot be applied.
The non-perturbative wave function scaling at the breakdown radius
does not imply however the failure of the perturbative treatment of OPE.
The reason is that the non-perturbative contribution (i.e. from
$r < 1\,{\rm fm}$) to the phase shifts is particularly small, 
as can be explicitly checked from applying the methods of
Ref.~\cite{PavonValderrama:2007nu} to this case, leading to
$$
\frac{d\,\delta^{(0)}_{{}^3P_1}(k; r_c)}{d r_c}
\sim k^3 \, r_c^{3/2} \, a_3^{1/2} \,
\,\exp{(-4 \sqrt{\frac{a_3}{r}})} \, .
$$
The vanishing exponential ensures that, once the non-perturbative regime
is achieved, the corresponding contributions to the total phase shift
will be negligible.
}.
For the particular case of the $^3P_1$ channel, this is not an unreasonable
prospect in view of the analysis of perturbative OPE
by Birse~\cite{Birse:2005um},
which suggests a critical momentum around $400\,{\rm MeV}$
above which the perturbative treatment will fail.
In contrast, the critical momentum for the attractive $^3P_0$ channel
is around $200\,{\rm MeV}$, requiring the non-perturbative treatment
of the OPE tensor force if we want to describe the phase shifts up
to $k \sim 2 m_{\pi}$.
In addition, the critical momenta quoted above were obtained in the 
$m_{\pi} \to 0$ limit.
For finite pion masses we expect the critical momenta to rise,
as the $1/r^3$ behaviour is strongly suppressed for $m_{\pi} r > 1$.
In the repulsive case, the potential barrier prevents the two nucleons to
explore in detail the $m_{\pi} r \sim 1$ region, where the singular
nature of the intermediate range tensor force manifests,
meaning that the critical momenta are expected
to rise significantly with respect to the 
$m_{\pi} \to 0$ limit.
On the contrary, in the attractive case, the two nucleons are driven to
the $m_{\pi} r \sim 1$ region, and thus the critical momentum
will only rise weakly.

However, there is still the problem of what happens if
the repulsive singular interaction is non-perturbative.
It is clear that the current understanding of the power counting
of repulsive singular potentials is incomplete.
Even though in the particular case of the $^3P_1$ partial wave
we can overcome the problem by invoking the perturbative
treatment of tensor OPE in this channel,
the issue may eventually have important consequences
for the power counting of coupled channels.
For the moment
we will simply assume that attractive and repulsive singular
interactions follow the same power counting,
unless we expect the singular potential to behave perturbatively.
It may be possible that the extension of the methods of
Long and Yang~\cite{Long:2011qx} from the $^3P_0$ to the $^3P_1$ channel 
will provide new insight into the power counting of
repulsive singular interactions.

\subsection{Coupled Channels}
\label{subsec:coupled}

For analyzing the perturbative renormalization of chiral TPE
in the coupled channel case, we employ the eigen
representation of the phase shifts~\cite{PhysRev.86.399}.
In this parametrization the DWBA formulas take their simplest form.
The eigen phase shifts are expanded according to the power counting
as follows
\begin{eqnarray}
\delta_{\alpha j}(k; r_c) &=& \sum^{\nu_{\rm max}}_{\nu = -1} \delta_{\alpha j}^{(\nu)}(k; r_c)  + {\mathcal O}(Q^{\nu_{\rm max} + 1}) \, , \\
\delta_{\beta j}(k; r_c) &=& \sum^{\nu_{\rm max}}_{\nu = -1} \delta_{\beta j}^{(\nu)}(k; r_c)  + {\mathcal O}(Q^{\nu_{\rm max} + 1}) \, , \\
\epsilon_{j}(k; r_c) &=& \sum^{\nu_{\rm max}}_{\nu = -1} \epsilon_j^{(\nu)}(k; r_c)  + {\mathcal O}(Q^{\nu_{\rm max} + 1}) \, ,
\end{eqnarray}
and the ${\rm LO}$ phase shifts are computed non-perturbatively
as explained in Appendix \ref{app:LO}.
The order $\nu$-th contribution to the eigen phase shifts is given by
\begin{eqnarray}
\label{eq:dnu_alpha_pert_unreg}
\frac{\delta_{\alpha j}^{(\nu)}(k; r_c)}{\sin^2{\delta_{\alpha j}^{(-1)}}} &=& - 
\frac{2\mu}{k^{2j-1}}\,{\mathcal{A}^{(-1)}_{\alpha j}}^2(k)\,
I_{\alpha \alpha j}^{(\nu)} (k; r_c) \, , \\ 
\frac{\delta_{\beta j}^{(\nu)}(k; r_c)}{\sin^2{\delta_{\beta j}^{(-1)}}} &=& - 
\frac{2\mu}{k^{2j+3}}\,{\mathcal{A}^{(-1)}_{\beta j}}^2(k)\,
I_{\beta \beta j}^{(\nu)} (k; r_c) \, , \\
\epsilon_j^{(\nu)}(k; r_c) &=& - \frac{2\mu}{k^{2j+1}}\,
\frac{{\mathcal{A}^{(-1)}_{\beta j}}(k)\,{\mathcal{A}^{(-1)}_{\alpha j}}(k)}
{\cot{\delta^{(-1)}_{\beta j}} -\cot{\delta^{(-1)}_{\alpha j}}}\,
I^{(\nu)}_{\beta \alpha j}(k; r_c) \, , \nonumber \\
\end{eqnarray}
where the perturbative integrals $I_{\alpha \alpha j}$, $I_{\beta \alpha j}$ and
$I_{\beta \beta j}$ are defined as
\begin{eqnarray}
I^{(\nu)}_{\rho \sigma j}(k; r_c) &=&
\int_{r_c}^{\infty}\,dr\,\Big[ V_{\rm aa}^{(\nu)}(r)\,
{u}_{k,\rho j}^{(-1)}\,(r)\,{u}_{k,\sigma j}^{(-1)}\,(r) + 
\nonumber \\ && V_{\rm ab}^{(\nu)}(r)\,\Big( 
{u}_{k,\rho j}^{(-1)}\,(r)\,{w}_{k,\sigma j}^{(-1)}\,(r) + \nonumber \\
&& \quad \quad \quad \quad
{w}_{k,\rho j}^{(-1)}\,(r)\,{u}_{k,\sigma j}^{(-1)}\,(r)
\Big) + 
\nonumber \\ && V_{\rm bb}^{(\nu)}(r)\,
{w}_{k,\rho j}^{(-1)}\,(r)\,{w}_{k,\sigma j}^{(-1)}\,(r) \Big] \, , 
\end{eqnarray}
with $\rho,\sigma = \alpha, \beta$; the subscripts ${\rm a}$, ${\rm b}$ 
in the potential, which take the values ${\rm a} = j-1$ and ${\rm b} = j+1$,
are used to denote the two angular momentum components of the potential
($l = j \pm 1$).
In turn, we employ $\alpha$, $\beta$ to label the two asymptotic solutions
at large distances, which behave as
\begin{eqnarray}
B_{\alpha j}^{(-1)}&&\,u^{(-1)}_{k,\alpha j}(r) \to \nonumber \\
&& \cos{\epsilon_j^{{(- \! 1)}}}\,
( \cot{\delta_{\alpha j}^{(- \! 1)}}\,\hat{j}_{j-1} (k r) - \hat{y}_{j-1} (k r) )
\, ,\nonumber \\
B_{\alpha j}^{(-1)}&&\,w^{(-1)}_{k,\alpha j}(r) \to \nonumber \\
&&\sin{\epsilon_j^{(- \! 1)}}\,
( \cot{\delta_{\alpha j}^{(- \! 1)}}\,\hat{j}_{j+1} (k r) - \hat{y}_{j+1} (k r) )
\, , \nonumber \\
\\
B_{\beta j}^{(-1)}&&\,u^{(-1)}_{k,\beta j}(r) \to \nonumber \\
&& -\sin{\epsilon_j^{(- \! 1)}}\,
( \cot{\delta_{\beta j}^{(- \! 1)}}\,\hat{j}_{j-1} (k r) - \hat{y}_{j-1} (k r) )
\, ,\nonumber \\
B_{\beta j}^{(-1)}&&\,w^{(-1)}_{k,\beta j}(r) \to \nonumber \\
&&\phantom{-}\cos{\epsilon_j^{(- \! 1)}}\,
( \cot{\delta_{\beta j}^{(- \! 1)}}\,\hat{j}_{j+1} (k r) - \hat{y}_{j+1} (k r) )
\, , \nonumber \\
\end{eqnarray}
where $B^{(- \! 1)}_{\alpha(\beta) j}$ are normalization factors,
which we define as
$k^{j-1}\,B^{(- \! 1)}_{\alpha j} = {\mathcal{A}^{(- \! 1)}_{\alpha j}}(k)$
and $k^{j+1}\,B^{(- \! 1)}_{\beta j} = {\mathcal{A}^{(- \! 1)}_{\beta j}}(k)$.
Finally, for obtaining the perturbative corrections to the phase shifts
in the nuclear bar representation~\cite{PhysRev.105.302},
which is the most widely used, we consider their relationship
with the eigen phases
\begin{eqnarray}
\bar{\delta}_{1j} + \bar{\delta}_{2j} &=&
{\delta}_{\alpha j} + {\delta}_{\beta j} \, , \\
\sin{( \bar{\delta}_{1j} - \bar{\delta}_{2j} )} &=&  
\frac{\tan{2\bar{\epsilon_j}}}{\tan{2\epsilon_j}} \, , \\
\sin{( {\delta}_{\alpha j} - {\delta}_{\beta j} )} &=& 
\frac{\sin{2\bar{\epsilon_j}}}{\sin{2\epsilon_j}} \, ,
\end{eqnarray}
and re-expand these relations according to the power counting.

As in the uncoupled case, the renormalizability of the coupled channel
phase shifts can be analyzed in terms of the $k^2$ expansion
of the perturbative integrals
\begin{eqnarray}
I^{(\nu)}_{\rho \sigma j}(k; r_c) = \sum_{n=0}^{\infty}\,
I^{(\nu)}_{2n, \rho \sigma j}(r_c) k^{2 n} \, ,
\end{eqnarray}
where only the first few terms in the expansion are divergent.
The properties of the $I^{(\nu)}_{2n, \rho \sigma j}$ terms are
in turn related to the $k^2$ expansion of the reduced wave functions,
which reads
\begin{eqnarray}
\begin{pmatrix}
u_{k,\rho j}(r) \\
w_{k,\rho j}(r)
\end{pmatrix} = 
\sum_{n=0}^{\infty}\,
\begin{pmatrix}
u_{2n, \rho j}(r) \\
w_{2n, \rho j}(r)
\end{pmatrix}\,k^{2 n} \, ,
\label{eq:uw-expansion}
\end{eqnarray}
where $u_{2n, \rho j}$ and $w_{2n, \rho j}$ behave as
\begin{eqnarray}
\begin{pmatrix}
u_{2n,\rho j}(r) \\
w_{2n,\rho j}(r)
\end{pmatrix} \sim
\begin{pmatrix}
r^{s_{{\rm a}\rho} + n t_{{\rm a}\rho}} \\
r^{s_{{\rm b}\rho} + n t_{{\rm b}\rho}}
\end{pmatrix} \, ,
\label{eq:uw-behaviour}
\end{eqnarray}
at short enough distances.
The specific values of $s_{l \rho}$ and $t_{l \rho}$ depend
on the details of the ${\rm LO}$ potential.
Owing to $t_{l \rho} \geq 2$, short distances are progressively more
suppressed, which in turn implies that the $I^{(\nu)}_{2n, \rho \sigma j}$
integrals will eventually converge for large enough $n$.

From the energy expansion of the reduced wave functions,
Eqs.~(\ref{eq:uw-expansion}) and ~(\ref{eq:uw-behaviour}),
we can see that the exact degree of divergence of
the perturbative integral depends on
the specific details of the channel and the phase shift under consideration.
However, two simplifications are applicable for the particular
case of the nuclear potential in chiral EFT.
The first is that all the components of the subleading order potential
diverge as $1/r^{3 + \nu}$, independently of the particular angular
momentum subchannels which are involved.
The second is that, for an iterated singular interaction at ${\rm LO}$,
the power law behaviour of the wave function at short distances
does not depend on the angular momentum or the $\alpha$ or $\beta$
nature of the scattering state:
in fact we have $s = s_{{\rm a}\alpha} = s_{{\rm b}\alpha} =
s_{{\rm a}\beta} = s_{{\rm b}\beta}$ and
$t = t_{{\rm a}\alpha} = t_{{\rm b}\alpha} =t_{{\rm a}\beta} = t_{{\rm b}\beta}$.
In this case, the situation is analogous to the uncoupled case and
the behaviour of the perturbative integrals is given by
\begin{eqnarray}
I^{(\nu)}_{2n, \rho \sigma}(r_c) \sim \int_{r_c} 
\frac{d\,r}{r^{3 + \nu - 2 s - n t}} \, ,
\end{eqnarray}
which diverges for $3 + \nu - 2 s - n t \leq 1$.

Independently of the previous simplifications, 
the essential point is that the perturbative integrals can be
renormalized by the addition of a certain number of counterterms.
As in the uncoupled channel case, the most explicit regularization method
is to include $n_{c(\rho \sigma)}$ free parameters, that is
\begin{eqnarray}
\label{eq:I_nu_modified_coupled}
\hat{I}^{(\nu)}_{\rho \sigma j}(k; r_c) = 
\sum_{n=0}^{n_{c (\rho \sigma)} - 1} \lambda_{\rho \sigma}^{(\nu)} k^{2n} +  
I^{(\nu)}_{\rho \sigma j}(k; r_c) \, , 
\end{eqnarray}
where $n_{c (\rho \sigma)}$ is the number of counterterms needed to regularize
the $\rho \sigma$ integral.
The total number of counterterms is simply
$n_{c} = n_{c (\alpha \alpha)} + n_{c (\alpha \beta)} + n_{c (\beta \beta)}$.

The relation between the $\lambda_{\rho \sigma}^{(\nu)}$ parameters and
the counterterms can be obtained by postulating an explicit
representation of the short range physics.
A convenient representation is
\begin{eqnarray}
\label{eq:VC-coupled}
V^{(\nu)}_{C,ll'}(r; r_c) &=&
i^{l-l'}\,\frac{f_l(r_c) f_{l'}(r_c)}{4\pi r_c^{2}} \nonumber \\
&\times&
\sum_{n=0}^{n_{c(ll')} - 1}
C^{(\nu)}_{2n,ll'}(r_c)\,k^{2n}\,\delta(r-r_c) \, ,
\end{eqnarray}
with $f_l(r_c) = \frac{(2l+1)!!}{r_c^l}$ and
where $n_{c(ll')}$ is the number of counterterms between the partial waves
$l$ and $l'$.
The relation between $n_{c(ll')}$ and $n_{c(\rho\sigma)}$ 
will be discussed below.
The momentum space representation of the the previous potential
in the $r_c \to 0$ limit takes the form
\begin{eqnarray}
\langle p | V^{(\nu)}_{C,l l'} | p' \rangle \to p^l {p'}^{l'}
\sum_{n=0}^{n_{c(ll')}- 1} C^{(\nu)}_{2n, ll'}(r_c)\,k^{2 n} \, , 
\end{eqnarray}
that is, they correspond with the usual representation
of energy-dependent counterterms.
If we have employed the short range normalization of
Eq.~(\ref{eq:uk-norm-short-prac}),
we obtain the relationship
\begin{eqnarray}
\label{eq:lambda-coupled-translation}
\lambda_{2n, \sigma \rho j}^{(\nu)} &=& 
\frac{f_a f_a}{4\pi r_c^{2}}\,C^{(\nu)}_{2n,aa}(r_c)
u_{\rho,0}(r_c)u_{\sigma,0}(r_c) - \nonumber \\ &&
\frac{f_a f_b}{4\pi r_c^{2}}\,C^{(\nu)}_{2n,ab}(r_c) \Big(
u_{\rho,0}(r_c) w_{\sigma,0}(r_c) + \nonumber \\ && 
\phantom{\frac{f_a f_b}{4\pi r_c^{2}}\,C^{(\nu)}_{2n,ab}(r_c)}
w_{\rho,0}(r_c) u_{\sigma,0}(r_c)
\Big) + \nonumber \\ &&
\frac{f_b f_b}{4\pi r_c^{2}}\,C^{(\nu)}_{2n,bb}(r_c)
w_{\rho,0}(r_c)w_{\sigma,0}(r_c) \, .
\end{eqnarray}
As can be seen, the previous representation of the short range physics
requires $n_{c(aa)} = n_{c(\alpha \alpha)}$, $n_{c(ab)} = n_{c(\alpha \beta)}$
and $n_{c(bb)} = n_{c(\beta \beta)}$.
This is however not a universal feature: most regulators only require
that the total number of counterterms remains the same, and
in general we should expect more counterterms in the lower partial
waves than with the delta-shell regularization.

A simplification that occurs in the coupled channel case is that
the ${\rm LO}$ interaction is always singular,
as a consequence of the tensor force being responsible of coupling channels
with different angular momenta.
Moreover the singularity structure of the tensor force implies the existence
of an attractive and a repulsive subchannel in all the coupled waves,
that is, we do not need to discuss the two cases separately.
The regular potential case will not be discussed either, but the results
are the expected: if we consider the regular solutions, we end up
with the standard Weinberg counting.
On the contrary, in the particular case of the $^3S_1-{}^3D_1$ partial wave,
if we consider the irregular solution for the s-wave,
the KSW counting appears.

\subsubsection{Singular Potential}

In coupled triplet channels, the tensor force piece of the OPE potential
behaves at short distances ($r \to 0$) as
\begin{eqnarray}
2\mu\,{\bf V}^{(-1)}_{\rm tensor}(r) \to \pm \,{\bf S}_j \frac{a_3}{r^3} \, ,
\end{eqnarray}
where ${\bf V}$ is a convenient matrix notation for the coupled channel
potential, that is
\begin{eqnarray}
{({\bf V}^{(-1)})}_{l l'} = V^{(\nu)}_{l l'} \, ,
\end{eqnarray}
with $l,l' = {\rm a},{\rm b} = j \pm 1$.
The matrix elements of the tensor operator, ${\bf S}^j$, reads
\begin{eqnarray}
{\bf S}_j = \frac{1}{2j+1}
\begin{pmatrix}
-2 (j -1) & 6\sqrt{j(j-1)} \\
6\sqrt{j(j-1)} & -2(j + 2)
\end{pmatrix}
\, .
\end{eqnarray}
Finally, $a_3$ is a length scale related to the strength of
the tensor force.

The interesting point for the regularization of the tensor force is
that, at short distances, it can be diagonalized by means of
the transformation
\begin{eqnarray}
2\mu\,{\bf R}_j\,{\bf V}^{(-1)}_{\rm tensor}\,{\bf R}^T_j = 
\pm\,\frac{a_3}{r^3}
\begin{pmatrix}
2 & 0 \\
0 & -4
\end{pmatrix} \, ,
\end{eqnarray}
which implies the existence of a repulsive and attractive eigenchannel.
Due to the $1/r^3$ singularity, the tensor force overcomes the centrifugal
barrier at short distances, and the reduced wave functions can be expressed
as a sum of the attractive and repulsive solutions.
In terms of the $k^2$ expansion of the reduced wave function,
Eq.~(\ref{eq:uw-expansion}),
the following behaviour is expected
\begin{eqnarray}
\begin{pmatrix}
u_{2n,\rho j}(r) \\
w_{2n,\rho j}(r)
\end{pmatrix} = r^{3/2 + 5 n / 2}\,
\begin{pmatrix}
g_{a \rho j}(x_A, x_R) \\
g_{b \rho j}(x_A, x_R)
\end{pmatrix} \, ,
\end{eqnarray}
where,  depending on the sign of the tensor force, we can either have
$x_A = 2\sqrt{2 a_3 / r}$, $x_R = 2\sqrt{4 a_3 / r}$ or 
$x_A = 2\sqrt{4 a_3 / r}$, $x_R = 2\sqrt{2 a_3 / r}$.
The functions $g_{a \rho j}$ and $g_{b \rho j}$ follow the general pattern
\begin{eqnarray}
g(x_A, x_R) = C_{S} \sin{x_A} +
C_{ C} \cos{x_A} +C_{R} e^{- x_R} \, ,
\end{eqnarray}
where we are only taking the regular solution for the repulsive eigenchannel.
The previous behaviour in turn implies that the repulsive solution plays
no significant role in the renormalization of the perturbative phase
shifts, as it only represents a negligible contribution to
the ${\rm LO}$ wave function at short enough distances.

From the previous form of the reduced wave functions, we have indeed
the simplifications which we expected as a consequence of iterating
a singular interaction at ${\rm LO}$, namely that
\begin{eqnarray}
s &=& s_{{\rm a}\alpha} = s_{{\rm b}\alpha} =
s_{{\rm a}\beta} = s_{{\rm b}\beta} = 3/4 \, , \\ 
t &=& t_{{\rm a}\alpha} = t_{{\rm b}\alpha} =t_{{\rm a}\beta} =
t_{{\rm b}\beta} = 5/2 \, .
\end{eqnarray}
If we compare with the s-wave regular solution of a regular potential
($s = 1$, $t = 2$),
the values above imply that the first perturbative correction to $C_0$
appears half an order earlier than expected, while the $C_{2n}$'s
happen $(n-1)/2$ orders later.
The finiteness condition reads
\begin{eqnarray}
\frac{5}{2} n_{c(\rho \sigma)} > \nu + \frac{1}{2} \, ,
\end{eqnarray}
which is independent of whether $\rho \sigma = \alpha \alpha$, 
$\alpha \beta$ or $\beta \beta$.
In terms of the scaling of the counterterms, the previous translates into
\begin{eqnarray}
C_{2n, l l'} \sim Q^{(5 n -1) /2} \, ,
\end{eqnarray}
that is, we obtain the expected scaling resulting from
the iteration of a $1/r^3$ potential.
However, due to the existence of the three different phase shifts,
the number of counterterms triple with respect to the uncoupled
channel case.
That is, at order $Q^{-1/2}$ we have a total of three counterterms,
at order $Q^{2}$ we reach six counterterms, and so on.

Even though the previous six counterterms guarantee the existence
of the $r_c \to 0$ limit of the ${\rm NLO}$ and ${\rm N^2LO}$
scattering amplitudes,
the existence of the repulsive eigenchannel in tensor OPE means
that finiteness may be probably assured with a smaller number
of contact operators~
\footnote{In particular, the two counterterms renormalizing
the attractive eigenchannel may be enough.
Of course, a formal proof will require to show that the divergences
in the perturbative integrals are correlated.
}.
The development of this argument requires to work in the basis
for which the OPE tensor force is diagonal
(i.e. the attractive-repulsive basis), instead of the usual partial wave basis.
This is non-trivial from the technical side and the problem remains of
which is the adequate power counting for the repulsive eigenchannel.
However, if we simply accept the proposal that the repulsive components
of the ${\rm LO}$ tensor force follow the same power counting
as the attractive ones~\cite{Birse:2005um},
the argument of finiteness becomes inconsequential from
the power counting point of view and
we end up with the aforementioned six counterterms
at ${\rm NLO}$/${\rm N^2LO}$.
This is the choice we are following for the coupled channels
in this work.

Alternatively, we can overcome the limitations in the understanding of
the power counting of repulsive singular interactions
if we suspect tensor OPE to be perturbative,
as happened in the $^3P_1$ partial wave,
in which case we can use the Weinberg counting.
There are two possibilities along this line: to work
(i) in the attractive-repulsive basis or (ii) in the partial wave basis.
The first case is interesting and leads to an exotic power counting~\footnote{
In particular we have $C_{2n, \rm AA} \sim Q^{(5n - 1)/2}$,
$C_{2n, \rm AR} \sim Q^{j - 5/4 + 2n}$ and
$C_{2n, \rm RR} \sim Q^{2 (j-1) + 2n}$,
where ${\rm A}$ and ${\rm R}$ refer to the attractive and repulsive
eigenchannel respectively.
It should be noted that the wave function in the repulsive eigenchannel
is a mixture of the $u_k$ ($l = j-1$) and $w_k$ ($l = j+1$)
wave functions, and consequently it behaves
as $l = j-1$ in terms of the Weinberg counting.
},
but is also difficult to implement,
as already commented in the previous paragraph,
and will not be considered in the present work.
The second case is much more straightforward and physically compelling
and will be analyzed in greater detail
in Sect. \ref{subsec:alternate}.

\subsection{Convergence of the Perturbative Series}
\label{subsec:convergence}

The convergence of the perturbative series can be understood
in two different senses,
namely with respect to the power counting expansion
and with respect to the cut-off.
The first kind of convergence is related with the EFT expansion parameter,
while the second is linked to the interpretation of
the cut-off within the EFT framework.

We do not provide here definitive arguments, but rather educated
guesses about these aspects of the theory.
In particular, we can establish some bounds on the breakdown scale and
the expansion parameter by considering the related EFT expansion
of the chiral potential in the first place.
From the power series of the potential, see Eq.~(\ref{eq:pot-expansion}),
it is apparent that the expansion parameter is formally 
$x_0 = {Q}/{\Lambda_0}$ 
(independently of which is the concrete value of $\Lambda_0$).
Naively, we should also expect a power series in terms of $x_0$ 
for the scattering amplitude.
However, we can argue on general grounds that the expansion parameter
of the amplitude will be bigger than that of the potential.
That is, we have
\begin{eqnarray}
T = \sum^{\nu_{\rm max}}_{\nu = -1} T^{(\nu)} +
\mathcal{O}\left( (\frac{Q}{\Lambda_1})^{\nu_{\rm max} + 1}\right)\, ,
\end{eqnarray}
instead of Eq.~(\ref{eq:amp-expansion}),
where the real expansion parameter is not $x_0 = {Q}/{\Lambda_0}$,
but rather $x_1 = {Q}/{\Lambda_1}$,
with $\Lambda_1$ ($< \Lambda_0$) the true breakdown scale corresponding
to the particular power counting under consideration.
The departure from the naive expectation $\Lambda_1 = \Lambda_0$
is explained in terms of the iteration of the subleading pieces
of the chiral potential, which can spoil the convergence of
the perturbative expansion of the scattering amplitude.

The previous idea can be illustrated with
the KSW counting~\cite{Kaplan:1998tg,Kaplan:1998we}:
in the singlet channel, the KSW breakdown scale is determined by considering
the effect of the iteration of OPE on the running of
the counterterms~\cite{Kaplan:1998we},
yielding $\Lambda_{1,s} = \Lambda_{\rm NN} \simeq 300\,{\rm MeV}$.
This value is much lower than the expected $\Lambda_0 \sim 0.5\,{\rm GeV}$.
The expansion is even less converging in the triplet channel:
the analysis of perturbative OPE made by Birse~\cite{Birse:2005um}
indicates a breakdown scale $\Lambda_{1,t} \simeq 100\,{\rm MeV}$,
in agreement with the results of Ref.~\cite{Fleming:1999ee}~\footnote{
In this regard, the KSW reformulation of Ref.~\cite{Beane:2008bt}
claims to have solved the convergence problems of the triplet channel.
}.
In this particular case we have a clear example in which the convergence
is severely limited as a consequence of
higher order perturbation theory~.
That is, the correct identification of the non-perturbative contributions
to the interaction 
is an essential ingredient for a convergent EFT formulation.

In the case of interest for this work,
the Nogga et al. counting~\cite{Nogga:2005hy},
the breakdown scale can be estimated by the deconstruction method of
Refs.~\cite{Birse:2007sx,Birse:2010jr,Ipson:2010ah},
in which the form of the short range interaction is 
determined by removing the pion exchange effects from
the phenomenological phase shifts.
This analysis suggests a breakdown scale of 
$\Lambda_{1,s} \simeq 270\,{\rm MeV}$
for the $^1S_0$ and $^1P_1$ singlet channels~\cite{Birse:2010jr,Ipson:2010ah}.
In the triplet channels, the deconstruction has only been performed
in the uncoupled p- and d-wave triplets ($^3P_0$, $^3P_1$ and $^3D_2$),
yielding $\Lambda_{1,t} \simeq 340\,{\rm MeV}$~\cite{Birse:2007sx}.
From the previous results, we expect the approximate expansion parameters
$x_{1,s} = {m_{\pi}}/{\Lambda_{1,s}} \simeq 0.5$ and
$x_{1,t} = {m_{\pi}}/{\Lambda_{1,t}} \simeq 0.4$
for the singlet and triplet channels respectively~\footnote{
It should be noticed that the deconstruction estimations for the
breakdown scale are computed for a cut-off radius of $r_c = 0.1\,{\rm fm}$.
However, as is explained in the next paragraph, EFT calculations
will not converge below a certain value of the coordinate space cut-off
probably around $0.5\,{\rm fm}$.
Although results computed below that point are not to be trusted,
the truth is that the deconstruction estimations are stable up to
$r_c = 0.8\,{\rm fm}$, which is within the values of the cut-off
which we regard as acceptable in this work.
In this regard, we expect the final value of the expansion parameter
to be similar to the estimations
based on Refs.~\cite{Birse:2007sx,Birse:2010jr,Ipson:2010ah}.
}.

The second type of convergence, that is, the convergence with respect
to the cut-off, can also be stated in terms of the potential
instead of the scattering amplitude.
In this regard, the appearance of the $1/r^{3+\nu}$ behaviour
in the chiral expansion of the potential indicates that
the potential is not convergent at short enough distances.
If we consider the explicit expansion of the chiral potential
at distances below the pion Compton wavelength ($m_{\pi} r < 1$),
that is
\begin{eqnarray}
V_{\rm NN}(r) \propto \frac{1}{r^3}\,\sum^{\infty}_{\nu = 0}
\frac{1}{(\Lambda_0 r)^{\nu}} \, ,
\end{eqnarray}
this feature becomes apparent:
the Taylor expansion does not converge if $r < R_0 \sim 1/\Lambda_0$.
Within the framework of a non-perturbative treatment of the chiral potential,
this argument requires the coordinate space cut-off to be bigger
than the breakdown radius, $r_c > R_0$, which we naively
expect to lie in the $R_0 = 0.1-0.5\,{\rm fm}$ region.
Otherwise the scattering amplitudes will start to diverge at high enough orders,
independently of the number of counterterms included in the computations.
This in turn sets a limit on the value of the cut-off in perturbative
calculations.
Again, as a consequence of subleading order iterations, the bound of
the cut-off will become softer, $R_1 > R_0$. 

From this observation, a natural interpretation of the cut-off arises
within the context of approximation theory, as originally
advocated by Stevenson~\cite{PhysRevD.23.2916}.
While the full EFT scattering amplitude (computed at infinite order)
is independent of the cut-off in the regions where the potential
converges as a consequence of containing an infinite number of counterterms,
the truncated EFT scattering amplitude (computed at finite order)
is cut-off dependent.
However, as far as we have included the necessary counterterms
guaranteeing renormalizability~\footnote{This condition is required for
the difference between the full result and the calculation at $\nu$-th
order to formally scale as $Q^{\nu + 1}$, see also the related comments
at the end of Sect. \ref{sub:n2lo}.}, we are free to chose whatever
value of the cut-off that provides the better convergence
properties towards the full result
and realizes the particular power counting under consideration.
In this sense, the cut-off is just a parameter controlling the convergence
of the theory, as recently stated by Beane et al.~\cite{Beane:2008bt}.

\subsection{Power Counting}
\label{subsec:counting}

The power counting arising from the perturbative treatment of chiral TPE
is summarized in Table \ref{tab:counterterms}.
The total number of counterterms in the Nogga et al. counting 
at ${\rm NLO}$ and ${\rm N^2LO}$ ($n_c = 21$)
is certainly bigger than the corresponding one
in the dimensional Weinberg counting ($n_c = 9$).
The counterterm pattern is fundamentally very similar
to the power counting obtained by Birse using
RGA~\cite{Birse:2005um}.
However, there are two minor differences with respect to the results
of Ref.~\cite{Birse:2005um} that are worth commenting:
(i) the counting of the $^3S_1$ partial wave,
and (ii) the size of the $C_{2n}$ operators.

In the RGA of Ref.~\cite{Birse:2005um}, three counterterms (instead of two)
are predicted for the $^3S_1$ channel.
As explained in Ref.~\cite{Valderrama:2009ei}, this discrepancy arises
as a consequence of the naive extension of
the idea of trivial and non-trivial fixed points of
the RG equations~\cite{Birse:1998dk,Barford:2002je},
which is only valid for regular long-range potentials,
to the singular interaction case in Ref.~\cite{Birse:2005um}.
However, from the point of view of perturbative renormalizability,
the different fixed points stem from the power-law behaviour of
the ${\rm LO}$ wave functions at short distances.
In this regard, regular long-range potentials require the existence
of two fixed points as a consequence of the existence of a regular
and irregular solution.
That is, if we have a regular solution
the matrix element of the counterterm is
\begin{eqnarray}
\langle V_C \rangle_{\rm reg} =
C(r_c) \int dr \frac{u_k^2(r)}{4 \pi r^2} \delta (r-r_c)
\sim C(r_c) \, r_c^{2 l} \, , \nonumber \\
\end{eqnarray}
which requires $C(r_c)$ to scale as $r_c^{- 2 l}$ (i.e. $Q^{2 l}$)
for having a non-trivial effect.
On the other hand, for the irregular solution we have instead
\begin{eqnarray}
\langle V_C \rangle_{\rm irr} \sim C(r_c) \, r_c^{-2 l-2} \, ,
\end{eqnarray}
explaining the $Q^{-2l - 2}$ scaling.
However, in the particular case of attractive singular potentials,
all the short distance solutions of the wave function show
the same power law behaviour, and therefore there is only
one type of power counting in this case,
which is given by the $Q^{-1/2}$ scaling as can be trivially checked
by the previous procedure.

The second difference lies in the order of the energy-dependent
operators, $C_{2n}$, which is higher in the present formulation
($C_{2n} \sim Q^{(5 n - 1)/2}$) than in the RGA of Ref.~\cite{Birse:2005um}
($C_{2n} \sim Q^{(4 n - 1)/2}$).
A possible explanation is that the $k^2$ expansion of the reduced
wave function for the attractive triplets generates
a stronger power law short range suppression than
the corresponding one for a regular potential,
$r^{5/2}$ for each power of $k^2$ instead of $r^2$,
shifting the $C_{2n}$ operators to higher orders by a factor of $Q^{n/2}$. 
Taking into account the previous effect,
which may have been overlooked in the analysis of Ref.~\cite{Birse:2005um}
as a result of the presence of two cut-offs,
the $C_{2n}$ operators are demoted from
$Q^{(4 n - 1)/2}$ to $Q^{(5 n - 1)/2}$,
reconciling RGA and perturbative renormalizability.
However, the reconciliation is also possible if we consider 
energy-dependent contributions to the chiral potential,
that is
\begin{eqnarray}
V^{(\nu)}(r) \sim \frac{k^{2 m}}{r^{3 + \nu - 2 m}} \, ,
\end{eqnarray}
where $2 m \leq \nu$, $k = \sqrt{\mu\,E_{\rm lab}}$
is the center-of-mass momentum and $E_{\rm lab}$
the laboratory energy.
This contributions are not forbidden by power counting,
but do not occur in the present formulation which
employ energy-independent potentials~\footnote{
With the exception of the relativistic corrections to one pion exchange,
for which the energy-dependent representation is chosen.}.
If they are taken into account, the counterterms should be promoted
from $Q^{(5 n - 1)/2}$ to $Q^{(4 n - 1)/2}$ in the perturbative
framework, recovering the initial result from Ref.~\cite{Birse:2005um}.
Finally, a third explanation is provided by the possibility
that the $r^{5/2}$ suppression related to each subtraction
may be contaminated by the low energy scale $\Lambda_{\rm OPE}$.
In such a case, the missing scale would imply the reinterpretation
of $r^{5/2}$ as $r^{2}\,\sqrt{r \, \Lambda_{\rm OPE}}$,
thus enhancing the $C_{2n}$ counterterm by $Q^{-n/2}$.

In principle, the discussion is merely academical as it only affects
the counting of counterterms beyond the order considered
in the present work.
However, at ${\rm N^3LO}$ ($Q^4$) taking one option or
the other can substantially change the counting:
if we assume the scaling $C_{2n} \sim Q^{(5 n - 1)/2}$,
the third counterterm in triplet channels does not appear until
order $Q^{9/2}$, between ${\rm N^3LO}$ and ${\rm N^4LO}$.
On the contrary, if we follow Ref.~\cite{Birse:2005um}, the triplet channel
$C_4$ counterterms will enter at $Q^4$, increasing the total number
of free parameters from 27 to 35 at ${\rm N^3LO}$.
As the energy-independent representation of the finite range potentials
is the preferred one (and as far as the $\Lambda_{\rm OPE}$
contamination hypothesis has not been checked),
we advocate for the first option.
In this regard, the recent power counting analysis of the $^3P_0$ channel
by Long and Yang~\cite{Long:2011qx}
also prefers the $Q^2$ scaling (instead of $Q^{3/2}$) for the $C_2$ operator
of the attractive triplet, in agreement with our observations.

\section{The P- and D-Wave Phase Shifts}
\label{sec:results}

\begin{figure*}[htt!]
\begin{center}
\epsfig{figure=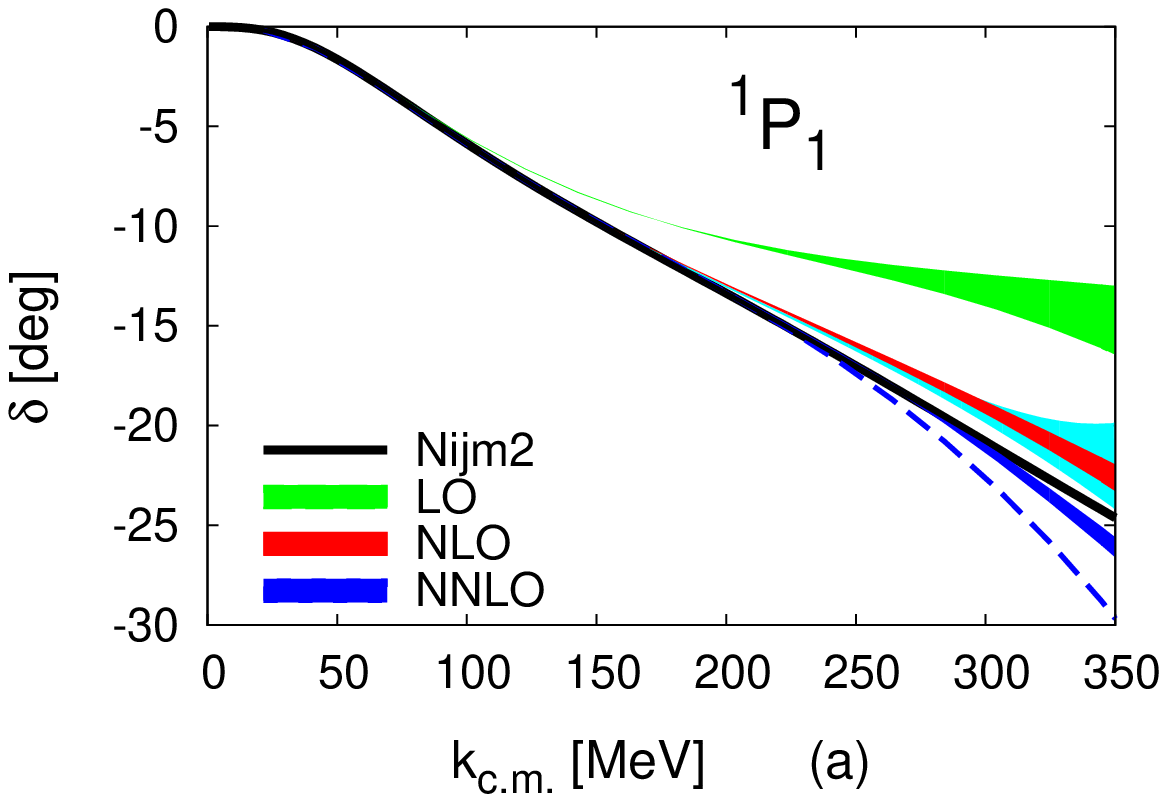, 
	height=4.75cm, width=8.0cm}
\epsfig{figure=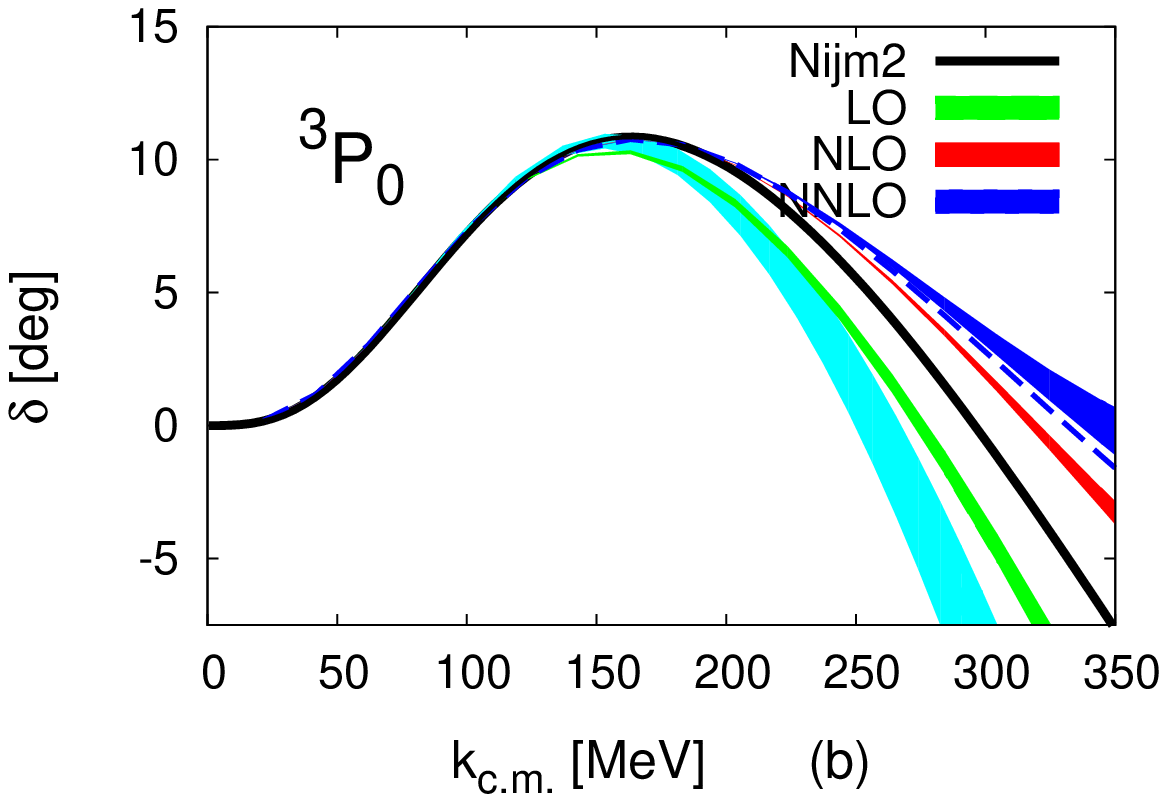, 
	height=4.75cm, width=8.0cm}
\epsfig{figure=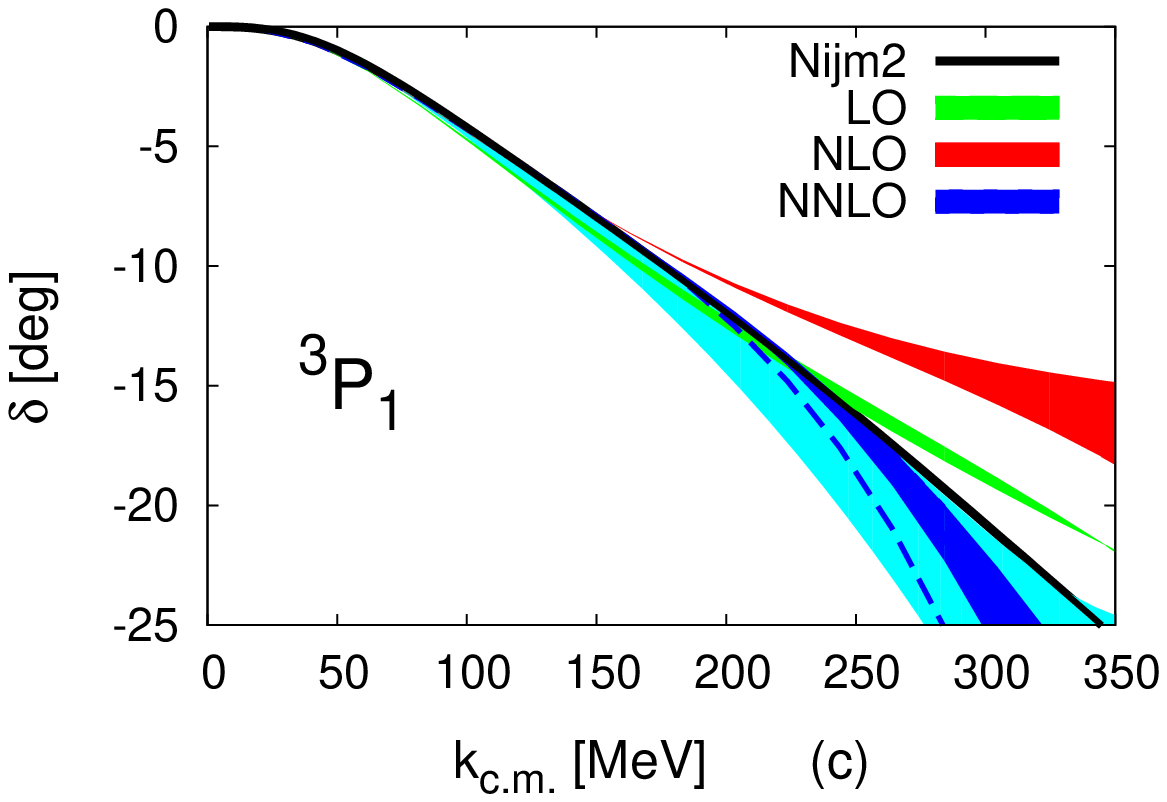, 
	height=4.75cm, width=8.0cm}
\epsfig{figure=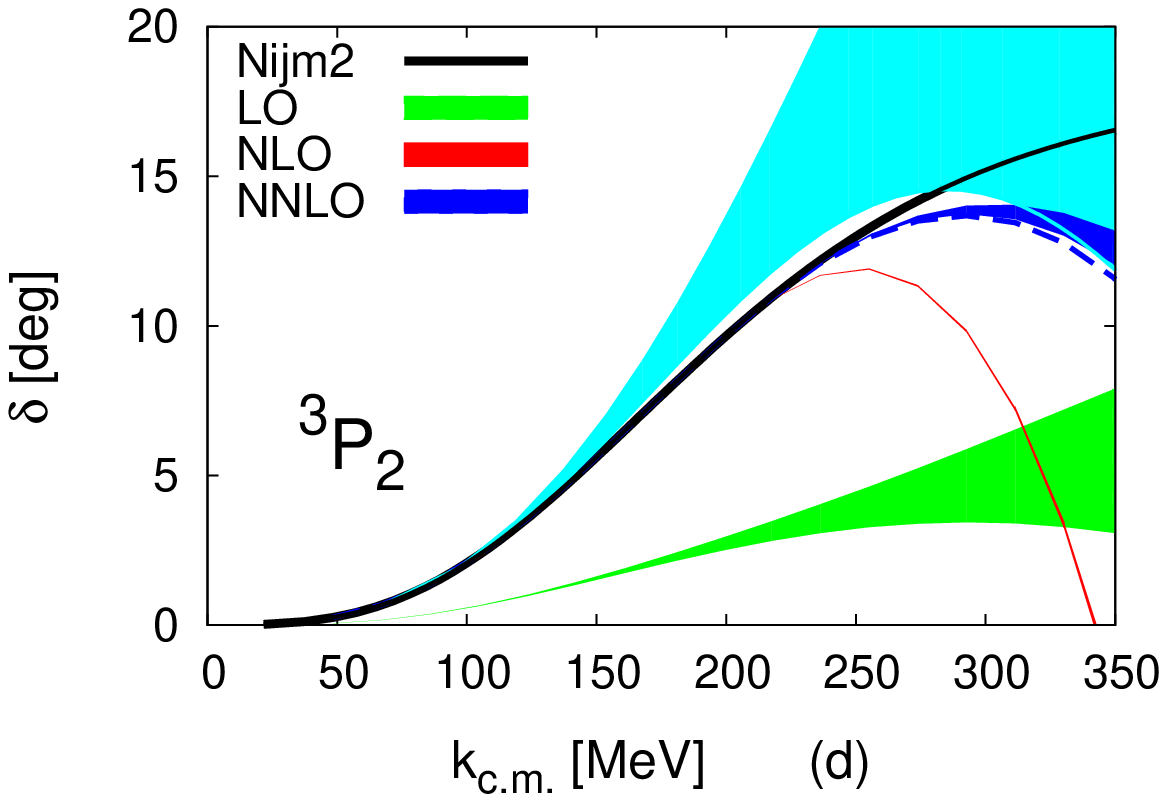, 
	height=4.75cm, width=8.0cm}
\epsfig{figure=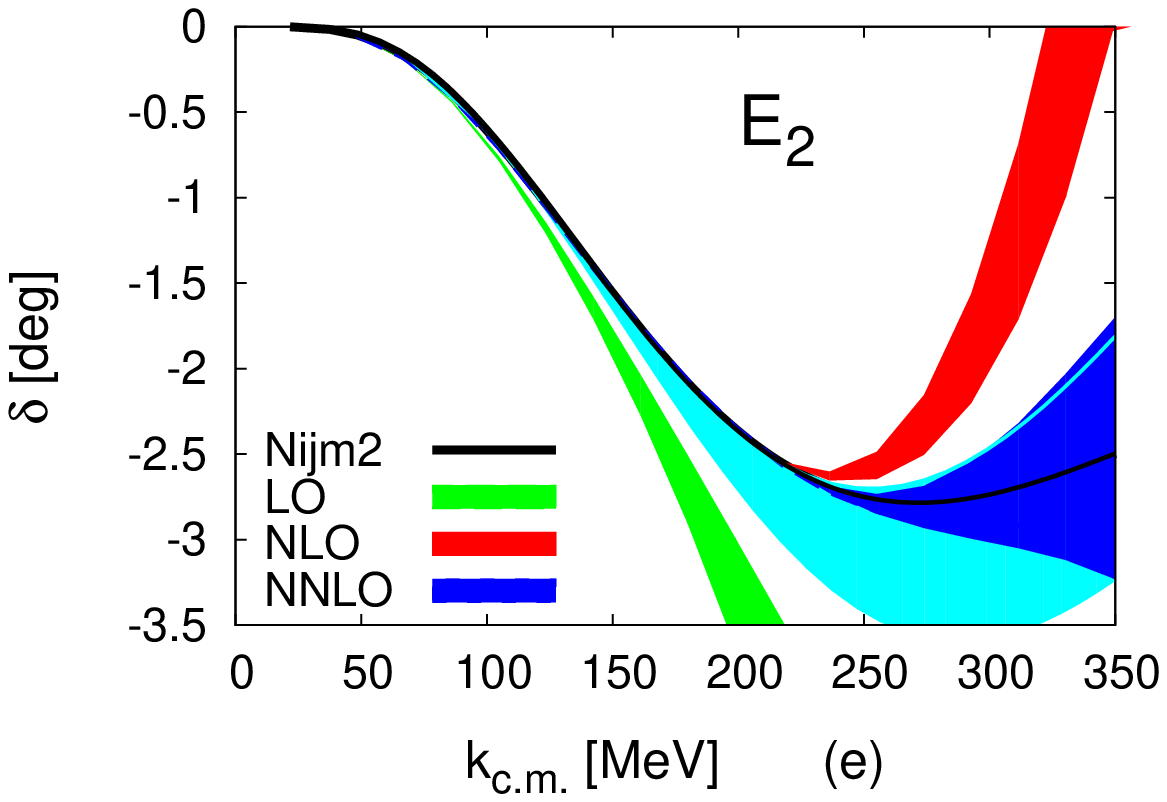, 
	height=4.75cm, width=8.0cm}
\epsfig{figure=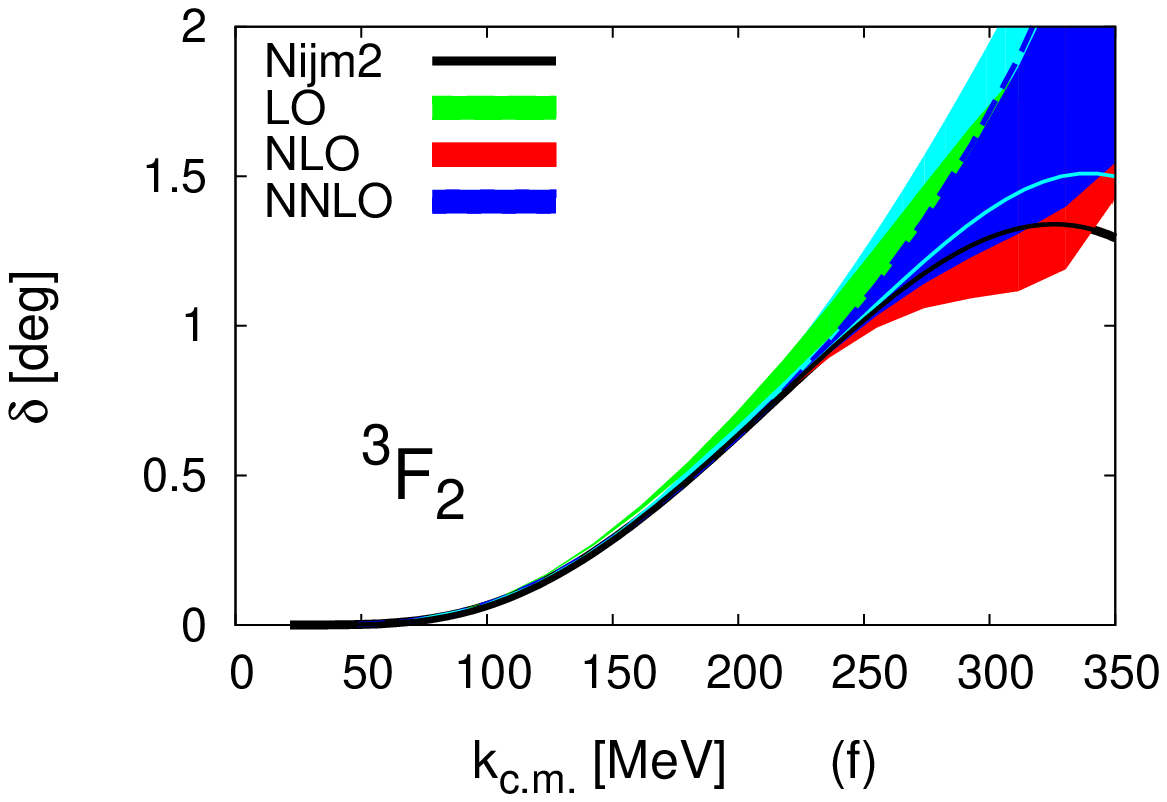, 
	height=4.75cm, width=8.0cm}
\end{center}
\caption{(Color online)
Phase shifts (nuclear bar) for the $^1P_1$, $^3P_0$, $^3P_1$ and
$^3P_2-{}^3F_2$ channels with the ${\rm LO}$ piece of
the chiral potential (OPE) fully iterated and the ${\rm NLO}$
and  ${\rm N^2LO}$ pieces (chiral TPE) treated perturbatively.
A ${\rm LO}$ counterterm in included in the attractive triplets
($^3P_0$, $^3P_2-{}^3F_2$) to remove the uncontrolled cut-off dependence.
The contact operators are used to fix the value of the scattering
lengths to $a_{^3P_0} = -2.7\,{\rm fm^3}$ and $a_{^3P_2} = -0.04\,{\rm fm^3}$.
The bands are generated by varying the cut-off radius in the region
$r_c = 0.6-0.9\,{\rm fm}$ and they are formally
a higher order effect.
However, as a result of the mild cut-off dependence of the results, 
the full uncertainty of the calculation at a given order
is expected to be much higher than the cut-off variation bands.
The light blue band corresponds to the ${\rm N^2LO}$ results
of Ref.~\cite{Epelbaum:2003xx} in the standard Weinberg counting.
The dashed dark blue line represents the ${\rm N^2LO}$ results for 
$r_c = 0.3\,{\rm fm}$.
}
\label{fig:p-waves}
\end{figure*}

In this section we compute the nucleon-nucleon p- and d-wave phase shifts
within the perturbative framework developed in the previous section,
that is, we treat the ${\rm LO}$ potential non-perturbatively and
include the subleading order corrections as perturbations.
For the counterterms, we follow the results of Table \ref{tab:counterterms}.
The details of the calculations are described in the following paragraphs.

The exact form of the chiral nucleon-nucleon potential in coordinate
space is taken from Ref.~\cite{Rentmeester:1999vw}.
We use the parameters $g_A = 1.26$, $f_{\pi} = 92.4\,{\rm MeV}$, 
$m_{\pi} = 138.03\,{\rm MeV}$, and $d_{18} = -0.97\,{\rm GeV}^2$.
The effect of the Goldberger-Treiman discrepancy ($d_{18}$)
in the chiral potential can be effectively taken into account
by changing the value of $g_A$ from $1.26$ to $1.29$
in the ${\rm LO}$ piece of the potential,
see Refs.~\cite{Epelbaum:2003gr,Epelbaum:2003xx} for details.
For the chiral couplings we use the values
$c_1 = -0.81\,{\rm GeV}^{-1}$, $c_3 = -3.40\,{\rm GeV}^{-1}$
and $c_4 = 3.40\,{\rm GeV}^{-1}$,
which are compatible with the determination of Ref.~\cite{Buettiker:1999ap}.
We include the relativistic corrections to OPE at ${\rm NLO}$
and the recoil corrections to TPE at ${\rm N^2LO}$.
Following Ref.~\cite{Rentmeester:1999vw}, the relativistic corrections to OPE
are included in an energy-dependent manner, that is 
\begin{eqnarray} 
\label{eq:OPE-rel}
V^{(2)}_{\rm OPE}(r) = -\frac{k^2}{2 M_N^2}\,V^{(0)}_{\rm OPE}(r) \, ,
\end{eqnarray}
where $k$ is the center of mass momentum and $M_N = 938\,{\rm MeV}$
is the nucleon mass.
It should be noted that the s-wave results of Ref.~\cite{Valderrama:2009ei}
did not include the relativistic correction to OPE.
However, explicit calculations show that the effects of the relativistic
corrections are fairly small: if we include the contribution from
Eq.~(\ref{eq:OPE-rel}), the s-wave phase shifts do not change
appreciably.
The same observation applies if we remove the recoil $1/M_N$ TPE corrections
from the results of Ref.~\cite{Valderrama:2009ei}.
In fact, if we consider the deconstruction estimation for the breakdown scale
of the theory, that is, $\Lambda_{1,s} \sim 300\,{\rm MeV}$ and
$\Lambda_{1,t} \sim 350\,{\rm MeV} $, we have the approximate
relationship
\begin{eqnarray}
\frac{Q}{M_N} \sim {\left( \frac{Q}{\Lambda_1} \right)}^2 \, ,
\end{eqnarray}
supporting the original convention of Weinberg
for the $Q/M_N$ terms~\cite{Weinberg:1991um}.

The ${\rm LO}$ phase shifts are computed non-perturbatively
by iterating the OPE potential to all orders and by adding
the counterterms required to achieve cut-off independence
at this order~\cite{Nogga:2005hy}.
This condition requires the inclusion of a counterterm in the
$^1S_0$, $^3S_1-{}^3D_1$,  $^3P_0$,  $^3P_2-{}^3F_2$ 
and $^3D_2$ partial waves at ${\rm LO}$.
In this work the counterterms are included in an implicit manner
by fixing the value of the scattering length
in the ${\rm LO}$ phase shifts
for the previous channels.
The technical details are explained in Appendix~\ref{app:LO}.
Following Ref.~\cite{Valderrama:2009ei}, we employ the cut-off window
$r_c = 0.6-0.9\,{\rm fm}$ for regularizing the phase shifts.
In addition, we show the results for $r_c = 0.3\,{\rm fm}$ for comparison,
which also serve as an informal check of the hard cut-off limit of
the scattering amplitudes.
However, contrary to what happens in the s-waves~\cite{Valderrama:2009ei},
the variation of the phase shifts in the previous cut-off window can be
hardly interpreted as the error band of the results at a given order.
The reason is that the cut-off dependence of the p- and d-wave phase shifts
is fairly small, sometimes negligible, for cut-off radii below
$r_c = 1.2\,{\rm fm}$.

The subleading order corrections to the phase shifts are computed by
making use of the perturbative formalism developed
in the previous section.
This formalism requires a total of 21 counterterms at ${\rm NLO}$/${\rm N^2LO}$,
distributed among the partial waves according to the results of
Table \ref{tab:counterterms}.
These counterterms, or equivalently the free parameters which modify
the perturbative integrals, see Eqs.~(\ref{eq:I_nu_modified}) and
(\ref{eq:I_nu_modified_coupled}),
are determined by fitting the ${\rm NLO}$ and ${\rm N^2LO}$
perturbative phase shifts in the $k = 100-200\,{\rm MeV}$ region
to the Nijmegen II phase shifts~\cite{Stoks:1994wp},
which are in turn equivalent to the Nijmegen
PWA~\cite{Stoks:1993tb}.

It should be noticed that the perturbative regularization techniques
employed in this work are specifically chosen to identify
divergences in the amplitudes rather than to optimize the phenomenology.
In fact, the regulator we employ can be considered to be the coordinate space
equivalent of the sharp regulator in momentum space, which is known
to be suboptimal from the phenomenological point of view.
Even with this proviso, the phase shifts we obtain are better than the
corresponding ones in the Weinberg scheme at the same order.
However, the amplitudes are clearly amenable to improvements.

\begin{figure*}[htt!]
\begin{center}
\epsfig{figure=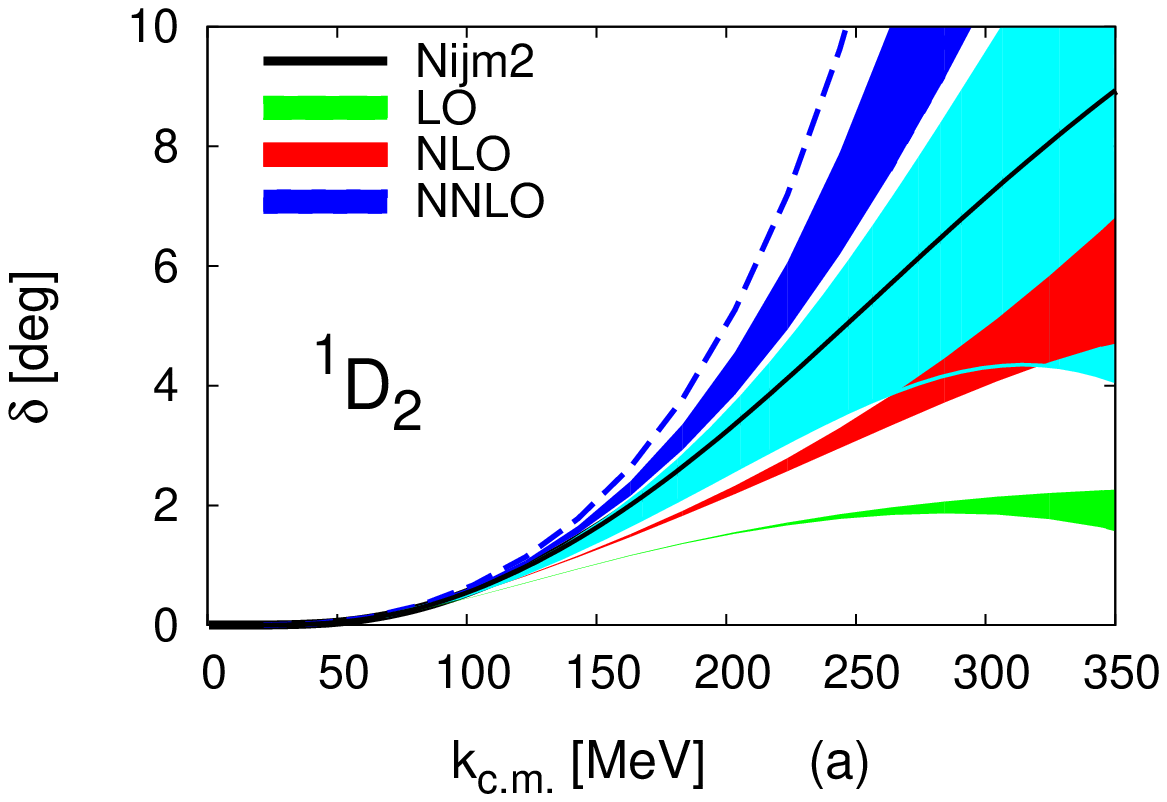, 
	height=4.75cm, width=8.0cm}
\epsfig{figure=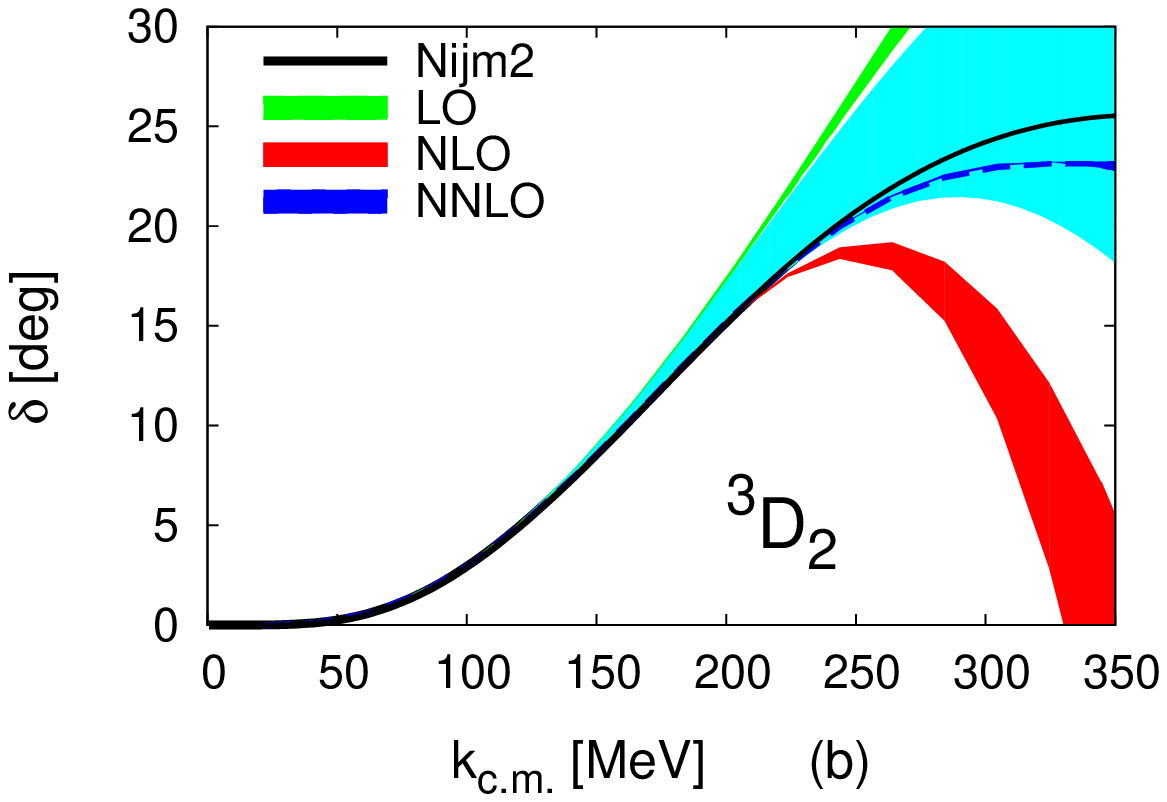, 
	height=4.75cm, width=8.0cm}
\epsfig{figure=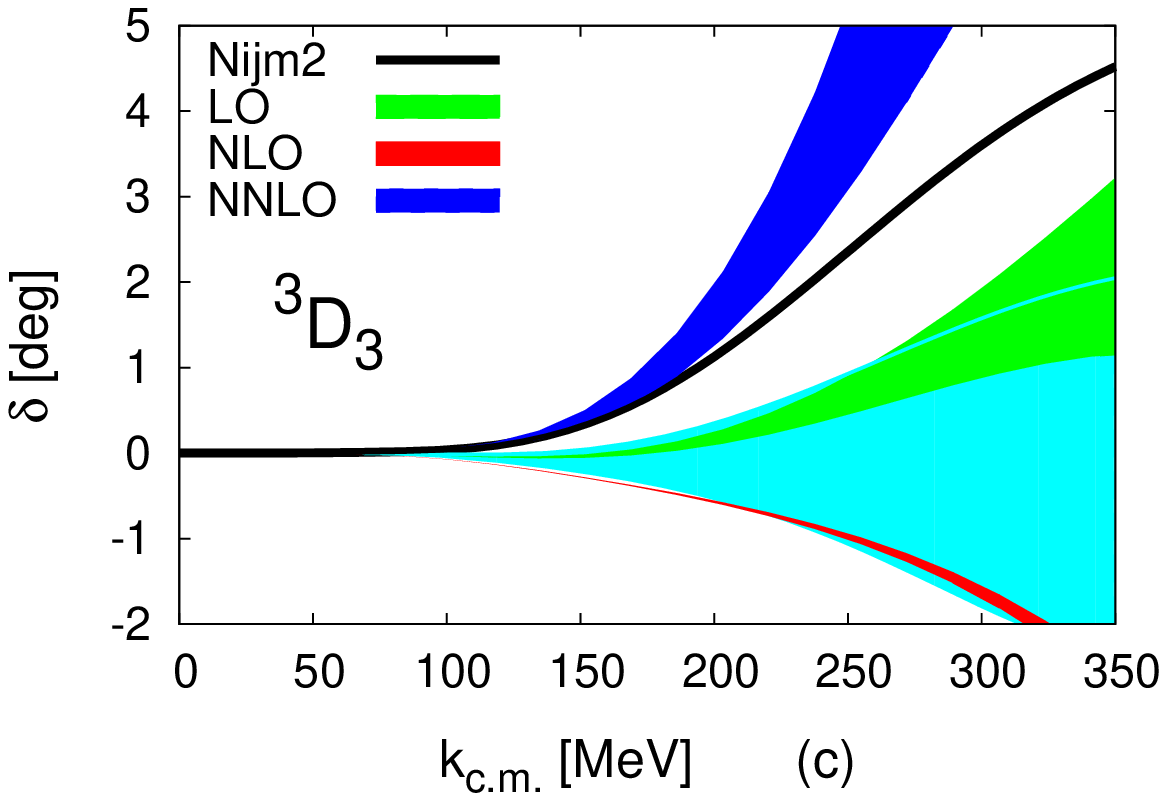, 
	height=4.75cm, width=8.0cm}
\epsfig{figure=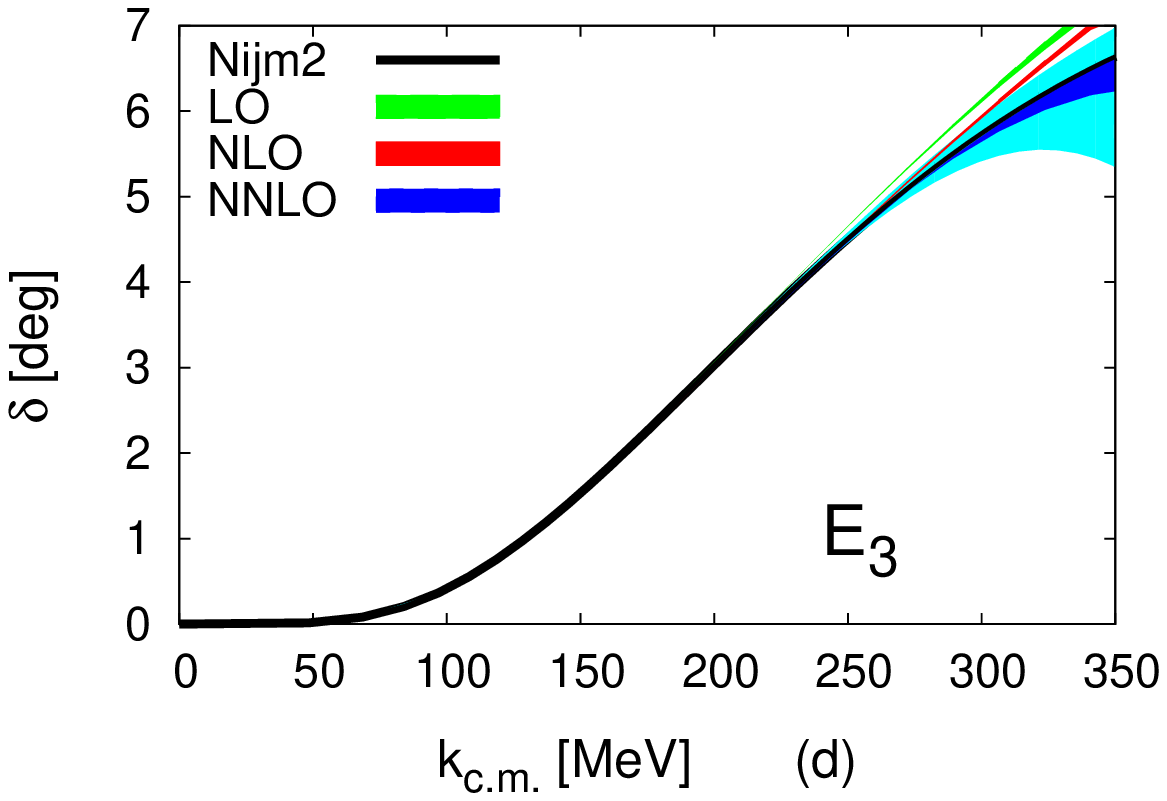, 
	height=4.75cm, width=8.0cm}
\epsfig{figure=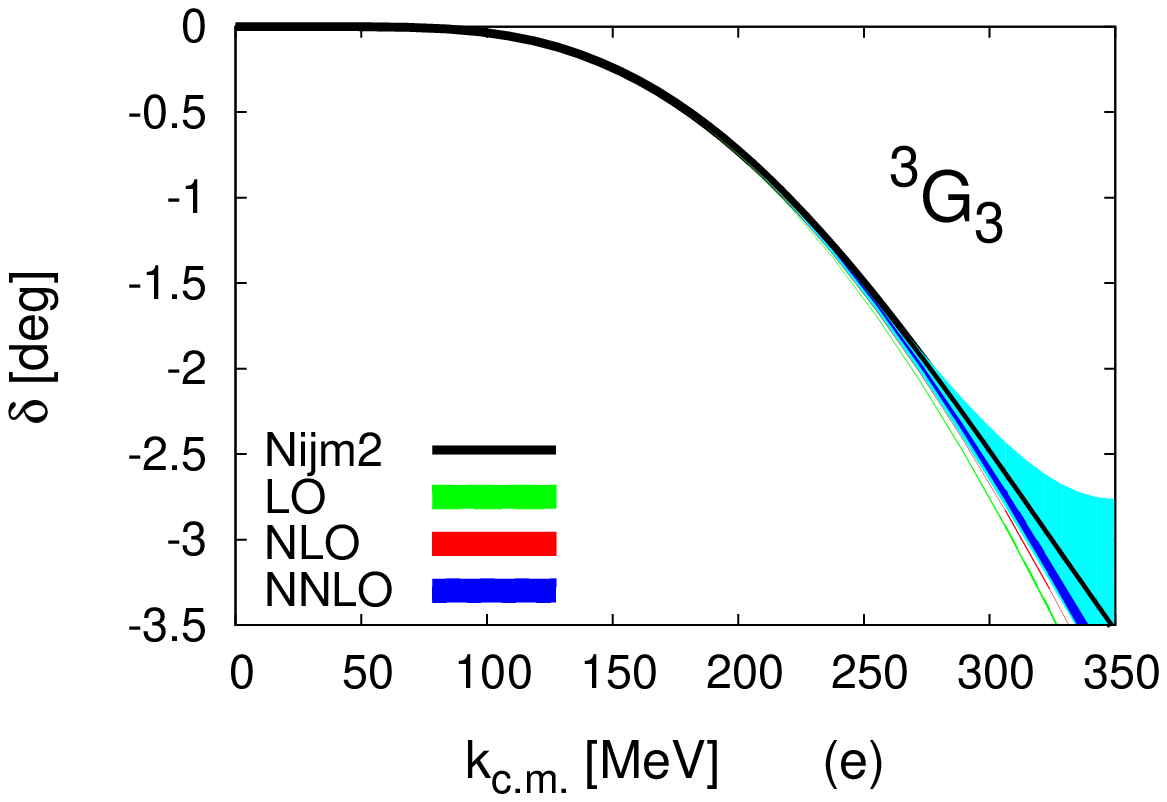, 
	height=4.75cm, width=8.0cm}
\end{center}
\caption{(Color online)
Phase shifts (nuclear bar) for the $^1D_2$, $^3D_2$ and $^3D_3-{}^3G_3$
channels, with the ${\rm LO}$ piece of the chiral potential (OPE)
fully iterated and the ${\rm NLO}$ and ${\rm N^2LO}$ pieces
(chiral TPE) treated perturbatively.
A ${\rm LO}$ counterterm is included in the $^3D_2$ channel,
as suggested in Ref.~\cite{Nogga:2005hy}, which is used to
fix the scattering length to the value $a_{^3D_2} = -7.4\,{\rm fm^5}$.
The bands are generated by varying the cut-off radius in the range
$r_c = 0.6-0.9\,{\rm fm}$ (see the related comments about
their interpretation in the previous figure and
in the main text).
The light blue band corresponds to the ${\rm N^2LO}$ results
of Ref.~\cite{Epelbaum:2003xx} in the standard Weinberg counting.
The dashed dark blue line represents the ${\rm N^2LO}$ results for 
$r_c = 0.3\,{\rm fm}$.
}
\label{fig:d-waves}
\end{figure*}

\subsection{P-waves}

The results for the p-wave phase shifts are shown in Fig.~(\ref{fig:p-waves}).
The renormalization of the ${\rm LO}$ phase shifts requires the inclusion of
a counterterm in each of the attractive triplets.
Therefore we fix the value of the scattering length in the $^3P_0$ and
$^3P_2-{}^3F_2$ channels to  $a_{^3P_0}= -2.7\,{\rm fm}^3$ and
$a_{^3P_2}= -0.04\,{\rm fm}^3$.
The previous values differ from the scattering lengths that can be obtained
from the Nijmegen II potential, namely $a_{^3P_0,{\rm Nijm}}= -2.468\,{\rm fm}^3$
and $a_{^3P_2,{\rm Nijm}}= -0.2844\,{\rm fm}^3$
according to Ref.~\cite{PavonValderrama:2004se}.
The ${\rm LO}$ deviations of the $^3P_0$ and $^3P_2-{}^3F_2$ scattering lengths
from the Nijmegen II values are however inconsequential:
the ${\rm LO}$ values should only be accurate
up to higher order corrections and there
is no need to use the exact value in the ${\rm LO}$ calculation.
The important point for the ${\rm LO}$ scattering lengths is
to provide a good starting point for the power counting expansion.
In addition, the mismatch between the peripheral effective range parameters
of the Nijmegen potentials and the preferred values of the chiral potential
is well known~\cite{PavonValderrama:2005uj,Yang:2009kx,Yang:2009pn,Valderrama:2010fb}.

In general, the description of the p-waves is better than in the
Weinberg counting, although this is a natural prospect once
we take into account the extra counterterms.
In this regard, the $^1P_1$ and $^3P_1$ wave are interesting in the sense
that they only contain the counterterms prescribed by Weinberg's original
counting.
Therefore they provide a more direct comparison with the Weinberg scheme,
and show that there is no significant drawback to treating TPE
perturbatively in these waves.
In fact, the results are slightly better than in the standard Weinberg
counting.

The phase shifts are nicely reproduced up to around a center-of-mass momentum
of $k_{\rm cm} = 300\,{\rm MeV}$, although there are signs of
convergence up to $k_{\rm cm} = 350\,{\rm MeV}$, a figure
compatible with deconstruction~\cite{Birse:2007sx}.
Contrary to the s-wave case, the value of the p-wave phase shifts
in the chosen cut-off window does not differ much from
those obtained when the cut-off is removed.
Consequently we expect the deconstruction estimations, which are obtained
for $r_c = 0.1\,{\rm fm}$ to work better
in the p-waves than in the s-waves.

\subsection{D-waves}

In Fig.~(\ref{fig:d-waves}) we show the results for the d-wave phase shifts.
In principle, the complete renormalization of the ${\rm LO}$ d-wave
scattering amplitudes requires the inclusion of a counterterm
in the two attractive triplets:
the $^3D_2$ and the $^3D_3-{}^3G_3$ partial waves.
However, we follow here the original suggestion of Ref.~\cite{Nogga:2005hy},
in which the counterterm is only (optionally) incorporated
in the $^3D_2$ channel.
The reason is that regulator dependence in the $^3D_3-{}^3G_3$ channel
is not apparent until relatively high values of the cut-off~\cite{Nogga:2005hy}.
Thus, we can safely obviate the counterterm,
at least in the ${\rm LO}$ calculation.
On the other hand, the full iteration of OPE in the $^3D_3-{}^3G_3$ channel
will require the inclusion of six additional counterterms, largely
reducing the predictive power of the theory.

In the $^3D_2$ channel we fix the value of the scattering length
to $a_{^3D_2} = -7.4\,{\rm fm}^5$, which basically coincides
with the Nijmegen II value
$a_{^3D_2,{\rm Nijm}} = -7.405\,{\rm fm}^5$~\cite{PavonValderrama:2004se}.
In the $^3D_3-{}^3G_3$ channel, we impose regular boundary conditions
for the ${\rm LO}$ amplitudes.
This procedure is of course non-renormalizable, as commented in the previous
paragraph, although the ${\rm LO}$ cut-off dependence does not appear
until we reach $r_c = 0.2\,{\rm fm}$.
However, the ${\rm NLO}$ and ${\rm N^2LO}$ cut-off dependence becomes manifest
at higher values of the cut-off, around $r_c = 0.5\,{\rm fm}$
in the ${\rm N^2LO}$ case.
That is, the present treatment of the $^3D_3-{}^3G_3$ phase shifts is
inconsistent.
This is also the reason why we do not show the $r_c = 0.3\,{\rm fm}$ results
for the $^3D_3-{}^3G_3$ partial wave in Fig.~(\ref{fig:d-waves}).

If we do not include the ${\rm LO}$ counterterm, the correct renormalization
of the $^3D_3-{}^3G_3$ channel should require the OPE potential to be of
order $Q^0$ instead of $Q^{-1}$.
Consequently, the consistent iteration of OPE requires the evaluation
of $(\nu+1)$-th order perturbation theory  to calculate
the $Q^{\nu}$ scattering amplitude,
a procedure which would generate renormalizable results.
On the contrary, the inconsistent full iteration of OPE at ${\rm LO}$
causes the amplitudes to eventually diverge as
\begin{eqnarray}
\int_{r_c} dr\,\frac{r^{3/2}}{r^{3+\nu}} \propto \frac{1}{r_c^{1/2+\nu}} \, ,
\end{eqnarray}
once the ${\rm LO}$ wave function attains the expected short range behaviour
for a solution of a $1/r^3$ potential.
Of course, the implicit assumption behind the present calculation of
the $^3D_3-{}^3G_3$ phase shifts is that the full-iterated (or inconsistent)
results are not going to be substantially different
from the partially-iterated (or consistent) ones,
provided the cut-off range is soft enough.
For the range of cut-offs we employ ($r_c = 0.6-0.9\,{\rm fm}$)
the divergent regime has still not been reached,
meaning that the current procedure probably represents
a fairly good approximation of a consistent result.
In any case, the elucidation of this aspect definitively calls
for a perturbative reanalysis of the $^3D_3-{}^3G_3$
partial wave in order to check this assumption.

\begin{table}
\begin{center}
\begin{tabular}{|c|c|c|c|c|}
\hline \hline
Partial wave & ${\rm LO}$ & ${\rm NLO}$ & ${\rm N^2LO}$ & ${\rm N^3LO}$ \\ 
\hline
$^3P_2-{}^3F_2$ & $1$ & $2$ & $3$ & $3$ \\ \hline
$^3D_3-{}^3G_3$ & $1$ & $2$ & $2$ & $3$ \\ \hline
All & 6 & 20 & 21 & 26 \\
\hline \hline
\end{tabular}
\end{center}
\caption{
Partial (and total) number of counterterms in the $^3P_2-{}^3F_2$ and
$^3D_3-{}^3G_3$ partial waves at ${\rm LO}$ ($Q^{-1}$),
${\rm NLO}$ ($Q^2$), ${\rm N^2LO}$ ($Q^3$) and ${\rm N^3LO}$ ($Q^4$)
assuming that (i) the iteration of the OPE potential is restricted
to the lower partial wave of the coupled channel ($^3P_2$, $^3D_3$)
and (ii) energy-dependent counterterms.
As in the previous case, for the ${\rm N^3LO}$ counterterms
see the related discussion of Sect.\ref{subsec:counting}.
 }\label{tab:counterterms-alternate}
\end{table}

It should be commented that
the d-wave (and to a lesser extend the f-wave) scattering amplitudes
have been traditionally a problem in nuclear EFT, especially at ${\rm N^2LO}$:
the excessive attractiveness of the central contribution to the chiral
potential at this order generates phase shifts which do not reproduce
the results of the partial wave analyses, as first noticed by Kaiser
et al.~\cite{Kaiser:1997mw,Kaiser:1998wa}.
In particular this poses an issue with respect to the convergence
of nuclear EFT in these waves: if the ${\rm N^2LO}$ results
are still considerably different from the PWA ones, then
we should still expect large corrections from higher
order contributions and, consequently, a slow convergence rate,
rendering the application of nuclear EFT impractical.
In this regard, Entem and Machleidt~\cite{Entem:2001cg} noticed
that the inclusion of a counterterm in each of the d-waves
in the ${\rm N^2LO}$ potential effectively solves the problem,
as the short range operators provide the necessary repulsion
for compensating the excess of attraction of the long range potential.
Curiously, this is not very dissimilar to the situation we encounter
within the present power counting scheme.
A different solution which is more compatible with the application of
the Weinberg counting is the use of spectral function regularization,
as advocated by Epelbaum et al.~\cite{Epelbaum:2003gr,Epelbaum:2003xx}.
In spectral function regularization a second cut-off is introduced to regularize
the momentum in the pion loops, a procedure which greatly reduces the
strength of the ${\rm N^2LO}$ potential, effectively solving the
convergence problem without the addition of new counterterms.
Of course, the most obvious solution is the use of softer cut-offs,
as happens for example with
the Nijmegen $\chi$PWA's~\cite{Rentmeester:1999vw,Rentmeester:2003mf}.

As we can see, the previous problem manifests moderately in the d-wave phase
shifts of Fig.~(\ref{fig:d-waves}).
In particular, it only affects the $^1D_2$ and $^3D_3$ phases,
although in a milder form than in Ref.~\cite{Kaiser:1997mw}.
In any case, the excess of attraction in the $^1D_2$ and $^3D_3$ channels
for the $r_c = 0.6-0.9\,{\rm fm}$ cut-off window is still under control,
and we do not expect to have convergence problems for
$k_{\rm cm} \leq 300\,{\rm MeV}$ at ${\rm N^3LO}$,
when the first counterterm is added.
However, it is clear that a softer cut-off window will probably improve
the convergence rate of these two partial waves without worsening
the other p- and d-wave phase shifts.
Finally, the $^3D_2$ partial wave is well-described and shows signs
of convergence up to $k_{\rm cm} = 350\,{\rm MeV}$, as expected
from deconstruction~\cite{Birse:2007sx}.

\subsection{Improved Power Counting Scheme for the Coupled Channels}
\label{subsec:alternate}

\begin{figure*}[htt!]
\begin{center}
\epsfig{figure=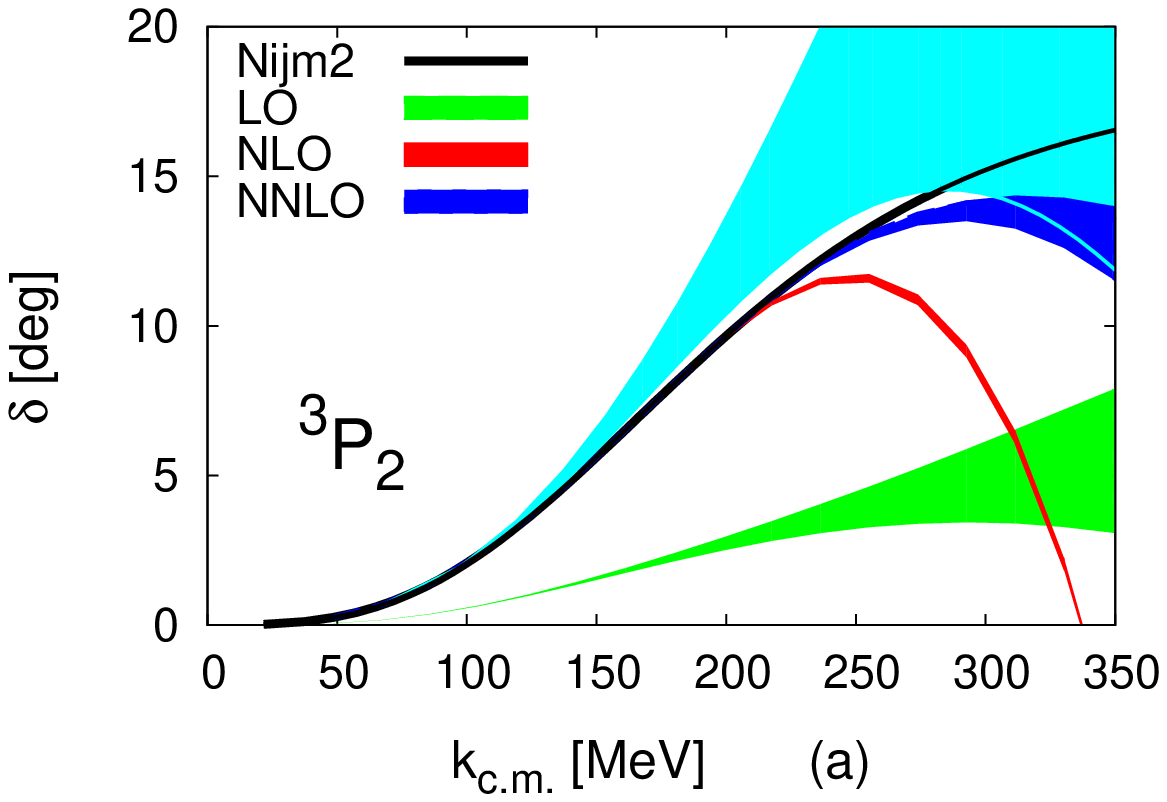, 
	height=4.75cm, width=8.0cm}
\epsfig{figure=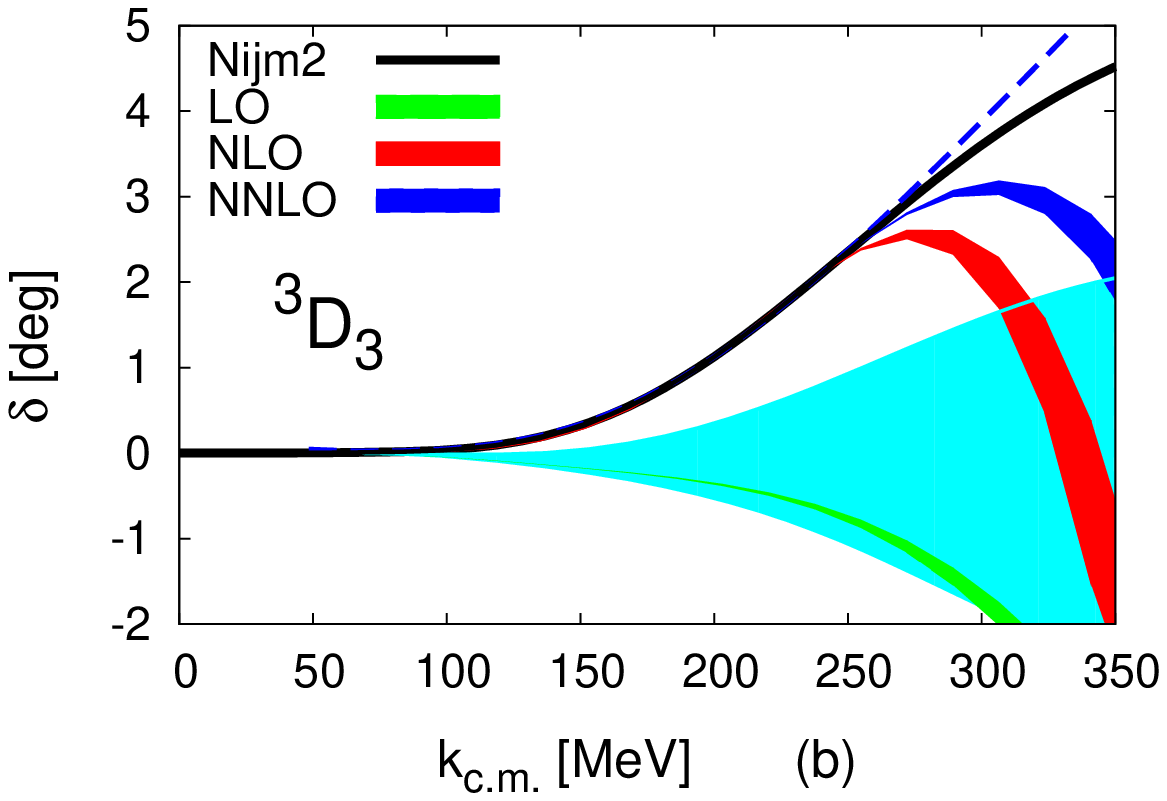, 
	height=4.75cm, width=8.0cm}
\epsfig{figure=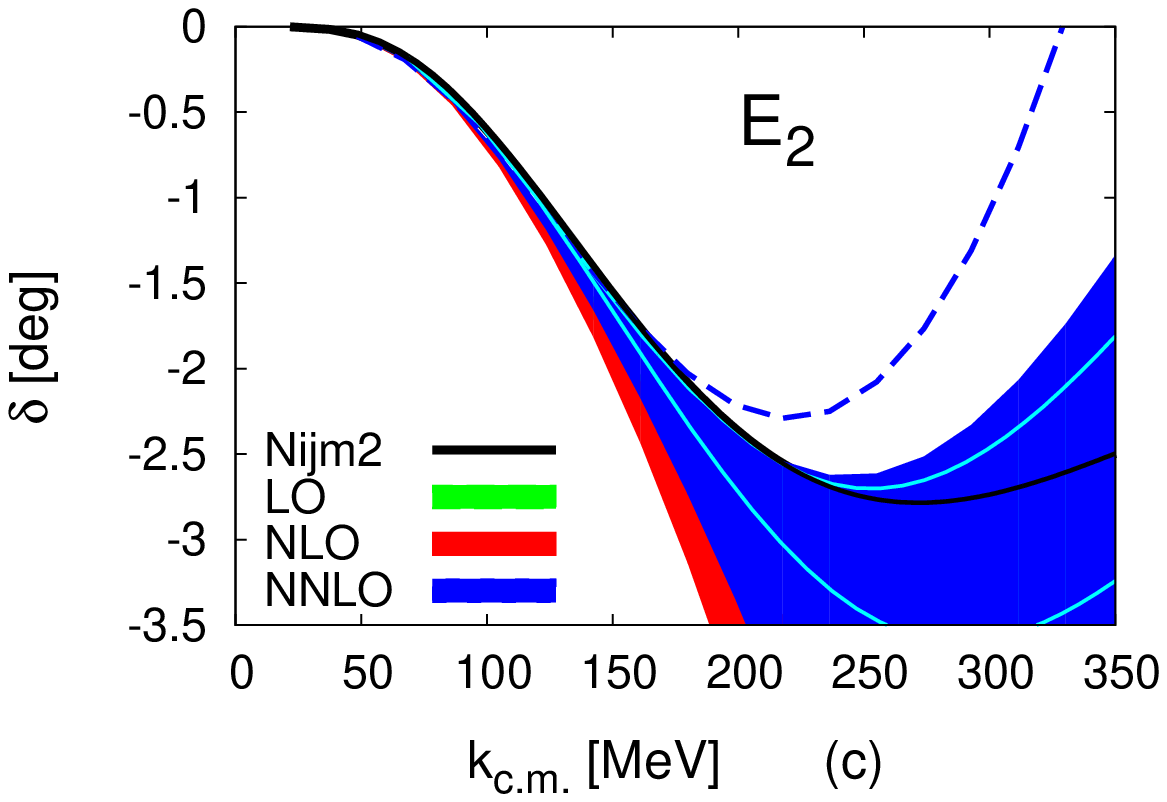, 
	height=4.75cm, width=8.0cm}
\epsfig{figure=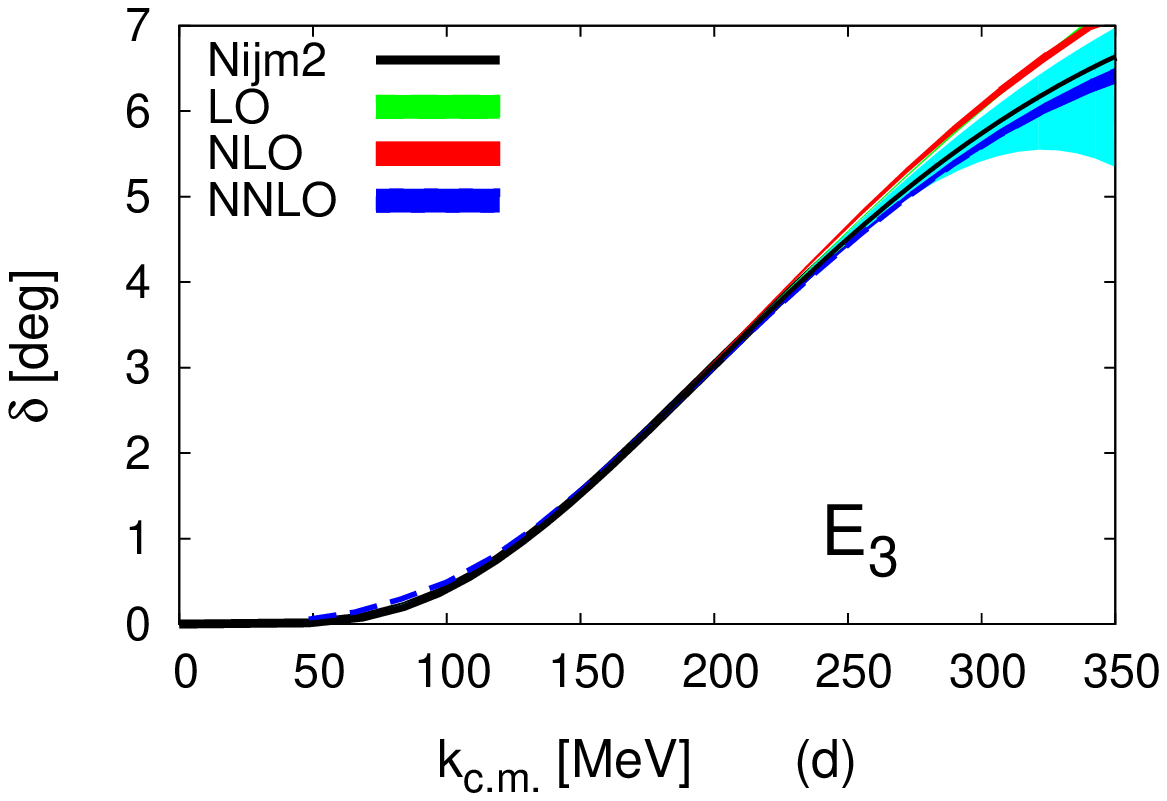, 
	height=4.75cm, width=8.0cm}
\epsfig{figure=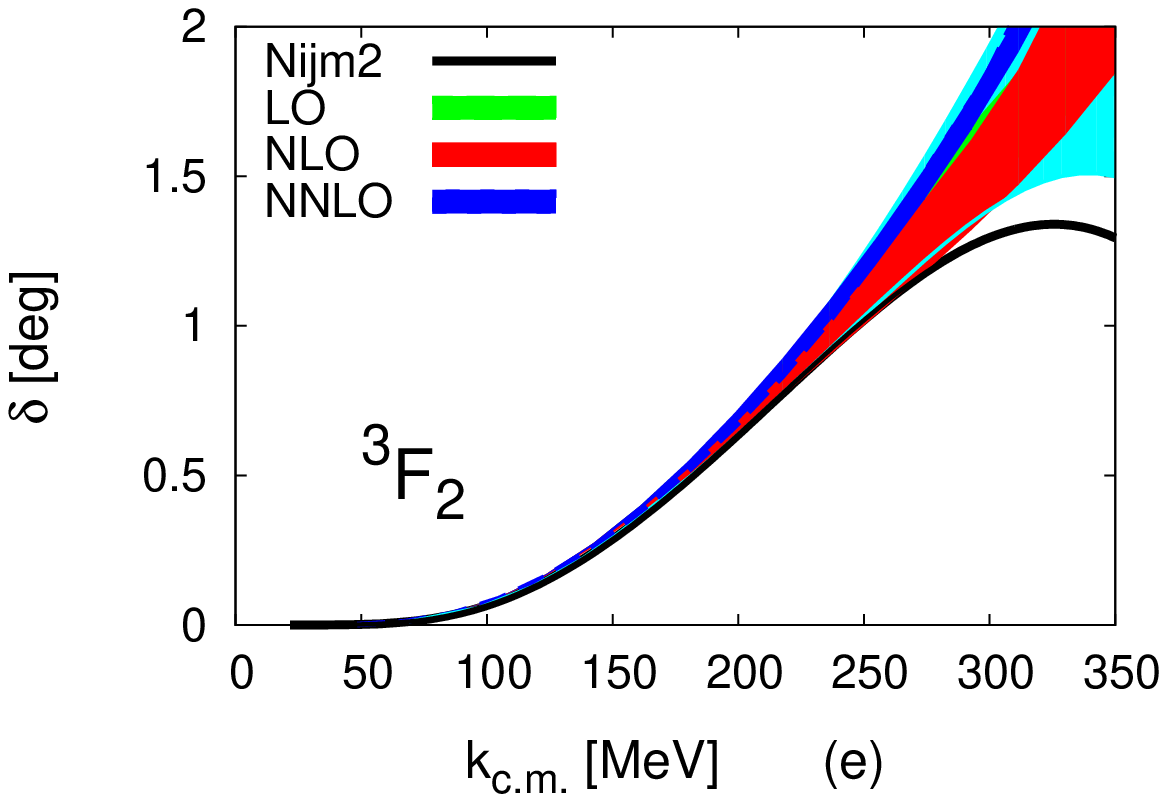, 
	height=4.75cm, width=8.0cm}
\epsfig{figure=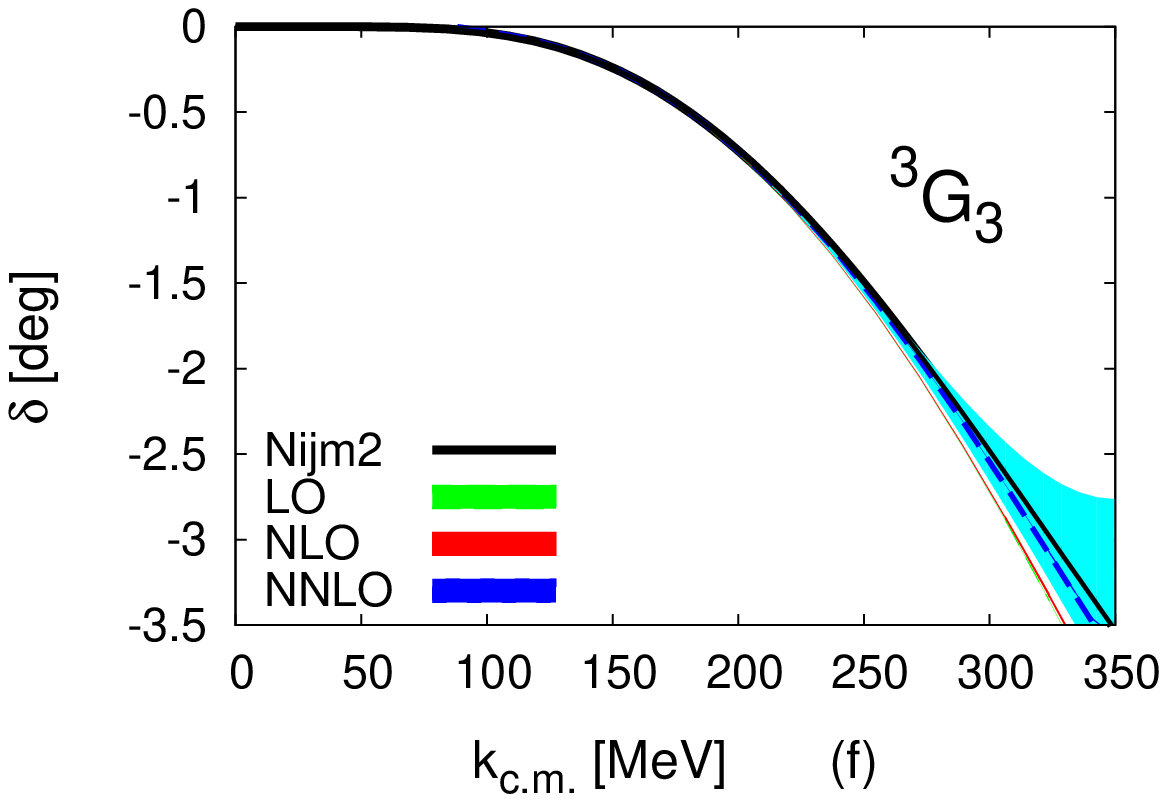, 
	height=4.75cm, width=8.0cm}
\end{center}
\caption{(Color online)
Phase shifts (nuclear bar) for the $^3P_2-{}^3F_2$ and
$^3D_3-{}^3G_3$ coupled channels in a power counting
scheme in which the OPE potential is fully iterated
only in the lower partial wave of the coupled channel,
while the other contributions are treated as perturbations.
The ${\rm LO}$ counterterm are employed to fix the value of
the scattering length to $a_{^3P_2} = -0.04\,{\rm fm^3}$ and
$a_{^3D_3} = -0.085\,{\rm fm^5}$.
The light blue band corresponds to the ${\rm N^2LO}$ results
of Ref.~\cite{Epelbaum:2003xx} in the standard Weinberg counting.
The dashed dark blue line represents the ${\rm N^2LO}$ results for 
$r_c = 0.3\,{\rm fm}$.
}
\label{fig:alt-waves}
\end{figure*}

A particularly problematic feature of iterating OPE
in the coupled triplet channels is
the associated requirement of including a total of six counterterms.
For the $^3S_1-{}^3D_1$ channel, this condition is not a significant drawback.
However, the counterterm proliferation caused by the full iteration of OPE 
starts to become worrisome in the $^3P_2-{}^3F_2$ channel
and excessive in the $^3D_3-{}^3G_3$ partial wave.
The reason for the six counterterm figure is that the iteration of OPE
in the coupled channels assumes the same short-range behaviour for the
two-linearly independent ${\rm LO}$ wave functions which correspond to
the  $\alpha$ and $\beta$ scattering states.
This justifies the necessity of two counterterms per phase: the scaling
of counterterms is identical independently of whether they only affect
the $^3P_2$ ($^3D_3$) partial wave, the $^3F_2$ ($^3G_3$),
or the admixture between them.
However, it is natural to expect that the $r^{3/4}$ wave function
behaviour will be first achieved in the $\alpha$
(rather than in the $\beta$) scattering state.

Along the previous lines we analyze here an alternative power counting scheme
for the peripheral coupled triplets which (i) reduces the number of
counterterms without significantly worsening the phenomenological
description of the phase shifts in the $^3P_2-{}^3F_2$ channel
and (ii) limits the counterterm proliferation
if we decide to iterate OPE in the $^3D_3-{}^3G_3$
partial wave.
The idea is to only iterate tensor OPE in the $\alpha$ scattering state 
of the coupled channel, that is, in the lower partial wave.
For the particular case of the $^3P_2-{}^3F_2$ channel what we mean is that,
while we have naively expected the full OPE potential
to be promoted to order $Q^{-1}$, that is
\begin{eqnarray}
\begin{pmatrix}
V^{(0)}_{pp} & V^{(0)}_{pf} \\
V^{(0)}_{fp} & V^{(0)}_{ff}
\end{pmatrix} \to
\begin{pmatrix}
V^{(-1)}_{pp} & V^{(-1)}_{pf} \\
V^{(-1)}_{fp} & V^{(-1)}_{ff}
\end{pmatrix} \, ,
\end{eqnarray}
it is much better to do the following promotion
\begin{eqnarray}
\begin{pmatrix}
V^{(0)}_{pp} & V^{(0)}_{pf} \\
V^{(0)}_{fp} & V^{(0)}_{ff}
\end{pmatrix} \to
\begin{pmatrix}
V^{(-1)}_{pp} & V^{(0)}_{pf} \\
V^{(0)}_{fp} & V^{(0)}_{ff}
\end{pmatrix}  \, 
\end{eqnarray}
in which only the diagonal p-wave piece of the potential is iterated.
In this scheme, we have one and two (p- to p-wave) counterterms at
orders $Q^{-1}$ and $Q^{2}$ respectively, while the first counterterm
mixing the p- and f-waves enters at order $Q^{2+3/4}$.
Taking into account the smallness of the ${\bar \epsilon}_2$ mixing parameter
and the ${\bar \delta}_{^3F_2}$ phase shift, the previous modification
does not seem an unreasonable prospect.
An advantage of this scheme is that the size of the matrix elements
of the two-pion exchange potential are reduced in the $pf$ and $ff$
channels, thus decreasing the size of the subleading contributions
with respect to the naive implementation of the counting
of Ref.~\cite{Nogga:2005hy}.
For the $^3D_3-{}^3G_3$ partial wave we propose the following promotion
\begin{eqnarray}
\begin{pmatrix}
V^{(0)}_{dd} & V^{(0)}_{dg} \\
V^{(0)}_{gd} & V^{(0)}_{gg}
\end{pmatrix} \to
\begin{pmatrix}
V^{(-1)}_{dd} & V^{(0)}_{dg} \\
V^{(0)}_{gd} & V^{(0)}_{gg}
\end{pmatrix}  \, ,
\end{eqnarray}
which implies one counterterm at ${\rm LO}$ for the ${dd}$ subchannel,
and two at ${\rm NLO}$ and ${\rm N^2LO}$.
The first counterterm mixing the d- and g-wave subchannels appears at
order $Q^{3+3/4}$, that is, between ${\rm N^2LO}$ and ${\rm N^3LO}$.
The power counting resulting from these modifications is summarized in 
Table \ref{tab:counterterms-alternate}.

The power counting proposed here can be explained
on the basis of a variation of
a well-known argument~\cite{Nogga:2005hy,Long:2011qx}
that explains the perturbative character of OPE in peripheral waves.
We analyze the behaviour of the partial wave projection
of the OPE potential at low momenta, which is given by
\begin{eqnarray}
\langle p | V_{\rm OPE, l l'} | p' \rangle
\sim \frac{c_{l l'}}{\mu\,\Lambda_{\rm OPE}}\,
\frac{p^l \, {p'}^{l'}}{m_{\pi}^{l+l'}} \, ,
\end{eqnarray}
where $\Lambda_{\rm OPE}$ is the scale governing the strength of OPE potential,
$l$ and $l'$ are the angular momenta of the partial waves
and $c_{l l'}$ a geometrical factor that approximately
behaves as $1 / (l! \, l'!)$~\cite{Nogga:2005hy}.
If $\Lambda_{\rm OPE}$ is considered to be a low energy scale,
the order of the OPE potential will be $Q^{-1}$
(instead of the naive $Q^0$).
However, owing to the far from perfect separation of scales,
the geometrical factor $c_{l l'}$ is eventually able to mimic
the effect of positive powers of $Q / \Lambda_0$
(or more properly $Q / \Lambda_1$, see Sect.\ref{subsec:convergence}
for details)~\footnote{There exists the possibility that a similar geometrical
suppression is taking place for repulsive tensor forces when treated
non-perturbatively, reconciling power counting with the naive expectation
that short range interactions should be less important
in the repulsive case compared to the attractive one.}.
In the nuclear EFT proposed in the present work, we expect a expansion
parameter of the order of $0.4$ for the triplet,
which means that for $l$ or $l' \geq l_{\rm crit}$
(with $l_{\rm crit} = 2$ or $3$) the angular momentum suppression
is equivalent to the demotion of OPE by at least one order. 
For higher partial waves the actual demotion of OPE will be even higher,
proportional to several orders in the chiral expansion,
thus explaining the fact that first order
perturbation theory for OPE works better
the more peripheral the partial wave~\cite{Kaiser:1997mw}.
For keeping matters simple, we have only considered the demotion
of OPE of by at most one order for $l \geq 3$.

Of course, the consistent implementation of the previous power counting
scheme requires the evaluation of perturbation theory up to third (fourth)
order at ${\rm NLO}$ (${\rm N^2LO}$).
However, owing to the computational difficulties related to the evaluation
of the perturbative series at high orders, we will fully iterate the OPE
potential as in previous cases.
Although this is inconsistent from the power counting point of view,
and probably even non-renormalizable,
we expect this simplified calculation to be a good approximation
of the true results in this improved counting scheme.
The regularization of the subleading order amplitudes is performed
by assuming a delta-shell parametrization of the counterterms,
see Eq.~(\ref{eq:VC-coupled}).
The corresponding $\lambda$-parameters to be added in the perturbative integrals
can be obtained by means of Eq.~(\ref{eq:lambda-coupled-translation}).
The counterterms are then fitted to the Nijmegen II phase shifts
in the $k_{\rm cm} = 100-200\,{\rm MeV}$ region.
The amplitudes generated by this particular regularization
procedure seem to be fairly cut-off independent:
no sign of divergent behaviour has been found for $r_c \geq 0.3\,{\rm fm}$.
Numerical limitations have prevented the exploration of the harder cut-off
region below $0.3\,{\rm fm}$.
Consequently we are unable to determine the eventual
renormalizability of the regularization we have employed~
\footnote{
Conversely, the direct inclusion of $\lambda$-parameters only
in the $I_{^3P_2}$ ($I_{^3D_3}$) and eventually
the $I_{E_2}$ perturbative integrals generate
divergences in all the integrals that
do not contain at least two $\lambda$-parameters.
}.

The results are shown in Fig.~(\ref{fig:alt-waves}).
The ${\rm LO}$ phase shifts have been regularized by fixing the
scattering length to $a_{^3P_2} = -0.04\,{\rm fm^3}$
and $a_{^3D_3} = -0.085\,{\rm fm^5}$.
In the $^3P_2-{}^3F_2$ channel the improved power counting scheme
yields similar results to the original one in Fig.~(\ref{fig:p-waves}),
but employing half the number of parameters.
The results for the $^3D_3-{}^3G_3$ have also improved, as expected
from the inclusion of two new counterterms.
The convergence properties are similar to other attractive triplets,
and the $r_c= 0.3\,{\rm fm}$ results do not significantly differ
from the $r_c = 0.6-0.9\,{\rm fm}$ ones.

\section{Discussion and Conclusions}
\label{sec:conclusions}

In the present work we have explored in detail the extension of
the power counting proposal of Nogga, Timmermans
and van Kolck~\cite{Nogga:2005hy}
to the subleading orders of the chiral expansion
up to ${\rm N^2LO}$ ($Q^3$).
We have chosen a perturbative regularization framework which
facilitates the identification of divergent contributions
and therefore the subsequent identification of the correct
power counting for the contact operators.
In particular, the present scheme fulfills the demanding constraints of
renormalizability and power counting without neglecting
the phenomenological aspects: the results are in fact
as good as, if not better than the corresponding ones
in the Weinberg counting at the same order.

We have determined the power counting of the contact operators,
resulting in a total of 21 counterterms for the s-, p- and d-waves
at ${\rm NLO}$ and ${\rm N^2LO}$, that is, twelve more than
in Weinberg dimensional counting at the same order.
The distribution of the contact operators largely agrees with
the related RGA by Birse~\cite{Birse:2005um},
but we also include several clarifications and improvements.
In the $^3P_1$ channel we explain the power counting disagreement
between RGA~\cite{Birse:2005um} and
the original exploration by Nogga et al.~\cite{Nogga:2005hy}.
We also propose modifications which enhance the consistency of the approach
and improve the counterterm distribution in the $^3P_2-{}^3F_2$
and $^3D_3-{}^3G_3$ channels.
However, a serious assessment of power counting in the previous cases
requires a perturbative reanalysis of the OPE potential up to fourth
order in the perturbative series.

In principle, the perturbative treatment of the subleading pieces of 
the chiral interaction raises the problem of a serious departure
with respect to the traditional approach to nuclear physics,
namely the idea of using a potential as the elementary computational
building block in few and many body calculations.
However, the divorce is merely apparent.
Perturbative renormalization should be understood as an analysis
tool of the formal aspects of the theory, rather than
as the adequate framework for final calculations.
The point is not to explicitly treat the subleading orders of
the chiral two-body forces perturbatively, but the fact that
they only represent a small contribution to physical
observables.

In this respect, a sensible approach is to construct chiral two- and
three-nucleon potentials that can be shown
(i) to contain the subleading pieces of the interaction approximately
as perturbations, therefore complying with power counting expectations, and
(ii) to be cut-off independent modulo higher order uncertainties
in a reasonable range of cut-offs.
Non-perturbative methods may fulfill these conditions provided
an adequate regulator function and cut-off window are employed,
as proposed in Ref.~\cite{Birse:2009my}.
In particular, the perturbative character of chiral TPE can always be
checked {\it a posteriori}.
If we consider the pioneering work of Ray, Ordo\~nez and
van Kolck~\cite{Ordonez:1993tn,Ordonez:1995rz}
as the first generation of chiral potentials,
and the phenomenologically successful potentials of
Entem and Machleidt~\cite{Entem:2003ft} and
Epelbaum, Gl\"ockle and Mei{\ss}ner~\cite{Epelbaum:2004fk}
as the second generation,
what we need is a third generation of chiral potentials
which overcomes the theoretical deficiencies of previous attempts.
This task will surely require much less work than the remaking
of all of nuclear physics in a perturbative fashion.

The prospects for the feasibility of this approach are indeed good.
In this regard, there exists the exciting possibility of reinterpreting
the Nijmegen $\chi$PWA99~\cite{Rentmeester:1999vw}
as an EFT calculation in the present power counting scheme.
In the chiral PWA of Ref.~\cite{Rentmeester:1999vw},
the authors regularize the chiral ${\rm N^2LO}$ potential
with a total of 23 (22) boundary condition at
the cut-off radius $r_c = 1.4\,(1.8)\,{\rm fm}$,
achieving a $\chi^2 / d.o.f. \simeq 1$
in the $E_{\rm lab} \leq 350\,{\rm MeV}$ region~\footnote{
Note that we are not taking into account the more recent $\chi$PWA03 of
Ref.\cite{Rentmeester:2003mf}, which claims to describe the nucleon-nucleon
scattering data up to $E_{\rm lab} = 500\,{\rm MeV}$ with a total of
33 boundary conditions at $r_c = 1.6\,{\rm fm}$.
The reason is that energy range of Ref.\cite{Rentmeester:2003mf}
extends well beyond the expected range of applicability of nuclear EFT.
}.
Taking into account the size of the cut-off radius they use,
chiral TPE is almost guaranteed to be a perturbation,
especially in view of the perturbative analysis
of Ref.~\cite{Shukla:2008sp}.
In the power counting advocated in this work, we employ a total of 21
parameters for the s-, p- and d-waves, a figure suspiciously similar
to Ref.~\cite{Rentmeester:1999vw}.
However, explicit calculations would be required to check
whether the $\chi$PWA99 can be considered a realization
of the counting proposal of Nogga et al.~\cite{Nogga:2005hy}.

\begin{acknowledgments}

I would like to thank E. Ruiz Arriola for discussions
and a critical and careful reading of the manuscript
and D. R. Entem for discussions.
This work was supported by the DGI under contracts FIS2006-03438
and FIS2008-01143, the Generalitat Valenciana contract PROMETEO/2009/0090,
the Spanish Ingenio-Consolider 2010 Program CPAN
(CSD2007-00042) and the EU Research Infrastructure Integrating Initiative
HadronPhysics2.

\end{acknowledgments}

\appendix

\section{Renormalization of the Leading Order Phase Shifts}
\label{app:LO}

In this appendix we explain the non-perturbative renormalization of
the ${\rm LO}$ scattering amplitudes and wave functions.
The appendix can be considered a review of the coordinate space renormalization
techniques of Refs.~\cite{PavonValderrama:2005gu,PavonValderrama:2005wv,PavonValderrama:2005uj}.
We consider first the uncoupled channel case and
then move to the coupled channel case.

\subsection{Uncoupled Channels}

We start by defining the reduced Schr\"odinger equation for the ${\rm LO}$
reduced wave function $\hat{u}^{(-1)}_{k,l}$
\begin{eqnarray}
\label{eq:schroe-LO-uncoupled}
-{\hat{u}^{(-1)}_{k,l}\,}'' + 
2\mu\,\left[ V^{(-1)}(r) + \frac{l(l+1)}{r^2} \right]\,\hat{u}^{(-1)}_{k,l} =
k^2 \hat{u}^{(-1)}_{k,l}\, , \nonumber \\
\end{eqnarray}
which we consider to be valid from from $r_c$ to $\infty$,
where $r_c$ is the cut-off radius.
With $k$ and $l$ we refer to the center of mass momentum and
the orbital angular momentum respectively,
and $V^{(-1)}(r)$ is the ${\rm LO}$ potential in configuration space.
The asymptotic ($r \to \infty$) normalization of the $\hat{u}_k$
wave function is given by
\begin{eqnarray}
\label{eq:uk-asymptotic}
\hat{u}^{(-1)}_{k,l}(r) \to \, k^l\,\left( 
\cot{\delta^{(-1)}_{l}}\,\hat{j}_l(k r) - \hat{y}_l(k r) \right)\, ,
\end{eqnarray}
where $\delta^{(-1)}_{l}(k; r_c)$ is the ${\rm LO}$ phase shifts.

The regularization and calculation of the phase shifts depend
on whether we include a counterterm at ${\rm LO}$ or not.
We will only consider in detail the first case,
as the second can be handled by standard means.
The addition of a counterterm is equivalent to the condition
of fixing the scattering length in the zero energy solution.
For this, we set the asymptotic form of the zero energy solution to
\begin{eqnarray}
\hat{u}^{(-1)}_{0,l}(r) \to \frac{(2l-1)!!}{r^l} - \frac{r^{l+1}}{(2l+1)!!}
\,\frac{1}{a_l^{(-1)}} \, ,
\end{eqnarray}
where $a_l^{(-1)}$ is the ${\rm LO}$ value of the scattering length,
which we fix to the desired value.
We integrate the zero energy solution downwards,
from $\infty$ to $r_c$ in Eq.~(\ref{eq:schroe-LO-uncoupled}).
Then, we assume the logarithmic derivatives of the zero and finite energy
solutions to be equal at $r = r_c$
\begin{eqnarray}
\frac{{\hat{u}_{k,l}^{(-1)}\,}'(r_c)}{\hat{u}_{k,l}^{(-1)}(r_c)} = 
\frac{{\hat{u}_{0,l}^{(-1)}\,}'(r_c)}{\hat{u}_{0,l}^{(-1)}(r_c)} \, ,
\end{eqnarray}
which effectively provides the initial integration conditions for obtaining
the finite energy solution at arbitrary radius $r$.
Finally we integrate upwards and obtain the phase shifts by comparing
the wave function at large radii with the asymptotic form of
Eq.~(\ref{eq:uk-asymptotic}).

However, for the previous wave functions to be useful in determining
the divergence of the perturbative integrals
we need to change the normalization.
In the normalization defined by Eq.~(\ref{eq:uk-asymptotic}),
the reduced wave function $\hat{u}^{(-1)}_{k,l}$ is not necessarily
energy independent at the cut-off radius,
making more difficult the identification of the required counterterms.
We therefore define a new normalization of the wave function
which is energy independent at $r = r_c$
\begin{eqnarray}
\hat{u}_{k,l}(r) = \mathcal{A}^{(-1)}_{l} (k; r_c) \, u_{k,l}(r) \, .
\end{eqnarray}
The normalization factor $\mathcal{A}^{(-1)}_{l} (k; r_c)$ can be uniquely
determined by requiring the additional condition
$\mathcal{A}^{(-1)}_{l} (0; r_c) = 1$,
in which case we have
\begin{eqnarray}
\mathcal{A}^{(-1)}_{l} (k; r_c) = \frac{\hat{u}_{k,l}(r_c)}{\hat{u}_{0,l}(r_c)}
\, .
\end{eqnarray}

\subsection{Coupled Channels}

For the coupled channel case the reduced Schr\"odinger equation reads
\begin{eqnarray}
\label{eq:schroe-LO-coupled}
-{\hat{u}^{(-1)}_{k,j}\,}'' + 
2\mu\,\left[ V_{aa}^{(-1)}(r) + \frac{(j-1) j}{r^2} \right]
&& \hat{u}^{(-1)}_{k,j}  \nonumber \\
+ 2\mu\,V_{ab}^{(-1)}(r)\,\hat{w}^{(-1)}_{k,j}
&=& k^2 \hat{u}^{(-1)}_{k,j}\, , \nonumber \\
\\
-{\hat{w}^{(-1)}_{k,j}\,}'' + 
2\mu\,\left[ V_{bb}^{(-1)}(r) + \frac{(j+1) (j+2)}{r^2} \right]
&& \hat{w}^{(-1)}_{k,j}  \nonumber \\
+ 2\mu\,V_{ba}^{(-1)}(r)\,\hat{u}^{(-1)}_{k,j}
&=& k^2 \hat{w}^{(-1)}_{k,j}\, , \nonumber \\
\end{eqnarray}
where $j$ is the total angular momentum and
$\hat{u}$ ($\hat{w}$) the $l = j-1$ ($l = j+1$) wave function.
For the ${\rm LO}$ potential we employ the $a$ ($b$) subscripts to
denote the $l = j-1$ ($l = j+1$) subchannel
in the potential matrix.
Of course, we assume the Schr\"odinger equation to be valid only
from $r = r_c$ to $\infty$.

For the coupled channel case we can define two linearly independent solutions
for $r \to \infty$, which are the $\alpha$ and $\beta$ scattering states.
In the Blatt-Biedenharn parametrization of the phase shifts~\cite{PhysRev.86.399} their asymptotic behaviour is given by
\begin{eqnarray}
\hat{u}^{(-1)}_{k,\alpha j}(r) &\to&   
k^{j-1}\,\cos{\epsilon_j^{{(- \! 1)}}}\,\nonumber\\
&\times& 
( \cot{\delta_{\alpha j}^{(- \! 1)}}\,\hat{j}_{j-1} (k r) - \hat{y}_{j-1} (k r) )
\, , \\
\hat{w}^{(-1)}_{k,\alpha j}(r) &\to&
k^{j-1}\,\sin{\epsilon_j^{(- \! 1)}}\,\nonumber\\
&\times&
( \cot{\delta_{\alpha j}^{(- \! 1)}}\,\hat{j}_{j+1} (k r) - \hat{y}_{j+1} (k r) )
\, , \\
\nonumber \\
\hat{u}^{(-1)}_{k,\beta j}(r) &\to&
- k^{j+1}\,\sin{\epsilon_j^{(- \! 1)}}\,  \nonumber \\
&\times&
( \cot{\delta_{\beta j}^{(- \! 1)}}\,\hat{j}_{j-1} (k r) - \hat{y}_{j-1} (k r) )
\, ,  \\
\hat{w}^{(-1)}_{k,\beta j}(r) &\to&
\phantom{-} k^{j+1}\,\cos{\epsilon_j^{(- \! 1)}}\, \nonumber \\
&\times&
( \cot{\delta_{\beta j}^{(- \! 1)}}\,\hat{j}_{j+1} (k r) - \hat{y}_{j+1} (k r) )
\, .
\end{eqnarray}

The renormalization of a singular potential in coupled channels
depends on the number of attractive or repulsive eigenchannels
at short distances~\cite{PavonValderrama:2005wv,PavonValderrama:2005uj}.
Here we will only consider in detail the kind of regularization which
appears when the ${\rm LO}$ interaction is
the OPE potential~\cite{PavonValderrama:2005gu}.
In such a case, we only need to fix one parameter to renormalize
the phase shifts.
This parameter is usually selected to be the scattering length associated
with the $\alpha$ phase shift, $\delta_{\alpha j}$,
but in principle there is no impediment
to fix any other low energy observable.
For fixing a scattering length we take into account
that the wave function of the zero energy
$\alpha$ and $\beta$ states behave asymptotically as
\begin{eqnarray}
\hat{u}^{(-1)}_{0,\alpha j}(r) &\to& \frac{(2j-3)!!}{r^{j-1}} - 
\frac{r^j}{(2j-1)!!}\,\frac{1}{a_{\alpha j}}
\, , \\  
\hat{w}^{(-1)}_{0,\alpha j}(r) &\to& \frac{(2j+1)!!}{r^{j+1}}\,e_j 
\, , \\
\nonumber \\
\hat{u}^{(-1)}_{0,\beta j}(r) &\to& \frac{r^{j}}{(2j-1)!!}\,
\frac{e_j}{a_{\beta j}} 
\, , \\  
\hat{w}^{(-1)}_{0,\beta j}(r) &\to& \frac{(2j+1)!!}{r^{j+1}}
- \frac{r^{j+2}}{(2j+3)!!}\,\frac{1}{a_{\beta j}}
\, , 
\end{eqnarray}
where $a_{\alpha j}$, $e_j$ and $a_{\beta j}$ are the scattering lengths
related to the $\delta_{\alpha j}$, $\epsilon_j$ and $\delta_{\beta j}$
phases (i.e. 
$\delta_{\alpha j} \to - a_{\alpha j} k^{2j-1}$, $\epsilon_j \to e_j k^2$,
$\delta_{\beta j} \to -a_{\beta j} k^{2j+3}$), 
see Ref.~\cite{PavonValderrama:2005uj} for details.
As previously suggested, we fix $a_{\alpha j}$ to the desired value,
leaving $e_j$ momentarily free, and integrate the zero energy
$\alpha$ state downwards from $r \to \infty$ to $r = r_c$.
At the cut-off radius we will impose boundary conditions
which will determine the initial integration conditions
for the finite energy $\alpha$ and $\beta$ scattering
states.

There are several possible regularization conditions
for the coupled channels~\cite{PavonValderrama:2005gu}.
In the present work, we will employ the following set
\begin{eqnarray}
{\hat u}_{k,\sigma}(r_c) &=& f_j \,{\hat w}_{k,\sigma}(r_c) \, , 
\label{eq:reg-condition-coupled}
\\
\frac{{\hat u}_{k,\sigma}(r_c)}
{f_j {\hat u}_{k,\sigma}'(r_c) + {\hat w}_{k,\sigma}'(r_c)} &=& 
\frac{{\hat u}_{0,\rho}(r_c)}
{f_j {\hat u}_{0,\rho}'(r_c) + {\hat w}_{0,\rho}'(r_c)} \, ,
\nonumber \\
\end{eqnarray}
where $\sigma, \rho = \alpha, \beta$ and with $f_j$ a parameter
which is set to $f_j = - \sqrt{\frac{j}{j+1}}$ for $j$ even
and $f_j = \sqrt{\frac{j+1}{j}}$ for $j$ odd.

The advantage of these particular regularization conditions is that
they facilitate the energy independent normalization of
the wave functions at short distances.
We define the new normalization as follows 
\begin{eqnarray}
\hat{u}^{(-1)}_{k,\sigma j} &=& \mathcal{A}^{(-1)}_{\sigma j}(k; r_c)\,
{u}^{(-1)}_{k,\sigma j} \, , \\
\hat{w}^{(-1)}_{k,\sigma j} &=& \mathcal{A}^{(-1)}_{\sigma j}(k; r_c)\,
{w}^{(-1)}_{k,\sigma j} \, ,
\end{eqnarray}
where $\sigma = \alpha, \beta$.
By requiring $\mathcal{A}^{(-1)}_{\sigma j}(0; r_c) = 1$,
we obtain the straightforward relationships
\begin{eqnarray}
\mathcal{A}^{(-1)}_{\sigma j}(k; r_c) = 
\frac{\hat{u}^{(-1)}_{k,\sigma j}(r_c)}{\hat{u}^{(-1)}_{0,\sigma j}(r_c)} =
\frac{\hat{w}^{(-1)}_{k,\sigma j}(r_c)}{\hat{w}^{(-1)}_{0,\sigma j}(r_c)} \, .
\end{eqnarray}
As can be seen, the regularization condition
given by Eq.~(\ref{eq:reg-condition-coupled})
assures that the normalization factor $\mathcal{A}^{(-1)}_{\sigma j}$
is well defined independently of whether we use the $\hat{u}$ or
$\hat{w}$ wave function for computing it.
Other regularization conditions do not guarantee this property,
thus generating a spurious energy dependence of one of the
components of the wave function at finite cut-off radii.


%

\end{document}